\documentclass[twocolumn]{aastex631}
\usepackage{aas_macros}
\usepackage{savesym}
\let\tablenum\relax
\usepackage{siunitx}
\usepackage[nameinlink,noabbrev]{cleveref}
\usepackage{hyperref}
\usepackage{xfrac}
\usepackage{color}
\usepackage{enumerate}

\newcommand{\lya}{Ly$\rm \alpha$}

\newcommand{\FLone}{SOL2010--10--16}
\newcommand{\FLtwo}{SOL2011--09--24}
\newcommand{\FLthree}{SOL2014--02--01}

\shorttitle{Observational Analysis of Lyman-alpha Emission in Equivalent Magnitude Solar Flares}
\shortauthors{Greatorex et al.}

\begin{document}

\title{Observational Analysis of Lyman-alpha Emission in Equivalent Magnitude Solar Flares}

\author[0000-0002-5302-0887]{Harry J. Greatorex}
\affiliation{Astrophysics Research Centre, School of Mathematics and Physics, Queen's University Belfast, University Road, Belfast, BT7 1NN, UK}

\author[0000-0001-5031-1892]{Ryan O. Milligan}
\affiliation{Astrophysics Research Centre, School of Mathematics and Physics, Queen's University Belfast, University Road, Belfast, BT7 1NN, UK}

\author[0000-0003-4372-7405]{Phillip C. Chamberlin}
\affiliation{Laboratory for Atmospheric and Space Physics, University of Colorado Boulder, Boulder, CO, 80309, USA}



\begin{abstract}


The chromospheric Lyman-alpha line of neutral hydrogen (\lya; 1216\AA) is the most intense emission line in the solar spectrum, yet until recently observations of flare-related \lya\ emission have been scarce. Here, we examine the relationship between nonthermal electrons accelerated during the impulsive phase of three M3 flares that were co-observed by RHESSI, GOES, and SDO, and the corresponding response of the chromosphere in \lya. Despite having identical X-ray magnitudes, these flares show significantly different \lya\ responses. The peak \lya\ enhancements above quiescent background for these flares were 1.5\%, 3.3\%, and 6.4\%. However, the predicted \lya\ enhancements from FISM2 were consistently $<$2.5\%. By comparing the properties of the nonthermal electrons derived from spectral analysis of hard X-ray observations, flares with a `harder' spectral index were found to produce a greater \lya\ enhancement. The percentage of nonthermal energy radiated by the \lya\ line during the impulsive phase was found to range from 2.0--7.9\%. Comparatively, the radiative losses in \ion{He}{2} (304\AA) were found to range from 0.6--1.4\% of the nonthermal energy while displaying enhancements above the background of 7.3--10.8\%. FISM2 was also found to underestimate the level of \ion{He}{2} emission in two out of the three flares. These results may have implications for space weather studies and modelling the response of the terrestrial atmosphere to changes in the solar irradiance, and will guide the interpretation of flare-related \lya\ observations that will become available during Solar Cycle 25.

\end{abstract}



\section{Introduction} \label{sec:intro}

During solar flares, a significant amount of magnetic free energy is released following the reconnection of opposing polarity magnetic fields. In the generally accepted standard model (CSHKP; \citealt{Carmichael1964CSHKP, Sturrock1966CSHKP, Hiryama1974CSHKP, KoppPneuman1976CSHKP}), liberated energy drives the acceleration of nonthermal electrons along newly restructured field lines towards the chromospheric footpoints of the flare loops. The majority of the energy deposited in the chromosphere during a solar flare is believed to be radiated by optical and extreme ultraviolet (EUV) emission \citep{Neidig1989WLF,Hudson2006WLFTRACE,Hudson1992YokkohWLF,Kleint2016ContinuumEnergy}. Measurements of the Total Solar Irradiance during solar flares by \cite{Woods2006TSIVariation} suggest that $\rm\sim$70\% of the total flare energy resides in wavelengths $>$270\AA, in good agreement with results from a superposed epoch analysis of solar flares conducted by \cite{Kretzschmar2010TSI} and \cite{Kretzschmar2011TSI}. Determining the radiated energy budget of solar flares, and how this energy is distributed across the spectrum, is an important step in constraining solar flare heating models (e.g. \citealt{Allred2005RADYN}).

In a multi-wavelength study of an X2.2 flare, \cite{Milligan2014EUVEnergy} were able to account for $\sim15\%$ of the total nonthermal energy deposited in the chromosphere through observations of various EUV/UV and optical lines and continua, leaving 85\% of the remaining energy unaccounted for. Of the available observations for this event, the \lya\ line of neutral hydrogen (\lya; 1216\AA) as measured by the \textit{Extreme Ultraviolet Variability Experiment} (EVE; \citealt{Woods2012EVE}) onboard the \textit{Solar Dynamics Observatory} (SDO; \citealt{Pesnell2012SDOMain}) was found to dominate the radiative losses in the chromosphere, amounting to 5--8\% of the total nonthermal energy as determined from hard X-ray (HXR) observations. 

\lya\ is the brightest emission line in the solar spectrum \citep{Curdt2001LyaBrightest} and during solar flares \lya\ emission is assumed to originate from the upper chromosphere/flare footpoints due to the relatively high abundance of hydrogen in these regions \citep{Chamberlin2018SolarStorm}. Enhancements in \lya\ irradiance during solar flares are generally marginal due to the intense background at this wavelength. \cite{Kretzschmar2012LyaProbaLYRA} reported a 0.6\%  enhancement in \lya\ above the background during an M2 flare measured with the \textit{Large-Yield Radiometer} onboard the \textit{Projects for Onboard Autonomy} mission (PROBA2/LYRA; \citealt{Dominique2012LYRA,Santandrea2013PROBA2}), and this was found to be comparable to \ion{He}{2} (304\AA) emission. This is consistent with the $<$1\% enhancements reported by \cite{Raulin2013LyaIonosphere} also using PROBA2/LYRA. By comparison, \cite{Woods2004Lya20Percent} found a peak enhancement of 20\% in the \lya\ line core and a factor of two increase in the wings for an X17 flare observed by \textit{Solar-Stellar Irradiance Comparison Experiment} onboard the \textit{Solar Radiation and Climate Experiment} (SORCE/SOLSTICE; \citealt{McClintock2005SOLSTICE}), whereas \cite{Breke1996X3Flare} found a 6\% enhancement in \lya\ for an X3 flare measured by the SOLSTICE instrument onboard the \textit{Upper Atmosphere Research Satellite} (UARS/SOLSTICE; \citealt{Rottman1993UARSPaper1, Rottman1993UARSPaper2}). 

In a statistical study of 477 M- and X-class flares observed in the E-channel of the \textit{Extreme Ultraviolet Sensor} onboard the \textit{Geostationary Operational Environmental Satellite} (GOES/EUVS-E; \citealt{Viereck2007EUVS,Evans2010EUVS}), \cite{Milligan2020MXFlares} found that typical increases in \lya\ were $<10\%$, with a maximum of approximately $30\%$. A follow-up study of several thousand flares using a superposed epoch analysis, revealed average increases of 0.18--0.35\% for B- and C- class flares, and 1--4\% for M- and X-class \citep{Milligan2021B&CClassFlares}.

The study of flare-induced changes in \lya\ irradiance is also important for space weather research. \lya\ is known to form and maintain the dayside D-region of Earth's ionosphere ($\rm \sim$60--90~km) through the photoionization of Nitric Oxide (NO; \citealt{Chubb1957SoundingRockets}), thus any fluctuations in solar \lya\ emission may have an impact on the dynamics of the terrestrial atmosphere. Despite this, observational studies of flare-induced \lya\ emission have been relatively scarce due to instrumental limitations, such as reduced duty cycles and insufficient sensitivity or cadence to capture changes in \lya\ irradiance on flare timescales \citep{Milligan2015EUVInvitedReview}. Of the studies into the terrestrial effects of flare radiation, a number have cited soft X-rays (SXRs) as the dominant driver of compositional changes in the ionosphere \citep{McRae2004D-RegionSXR,Kumar2018D-RegionSXR, Nina2018D-RegionSXR,Hayes2021Ionosphere}. \cite{Raulin2013LyaIonosphere} investigated the response of the D-region of the ionosphere to seven medium-sized flares, three of which had detectable \lya\ irradiance increases as observed by PROBA2/LYRA. Their analysis found that the measured Very Low Frequency (VLF) phase shift resulting from each flare in the sample showed no dependence on \lya\ radiation and concluded that the response of the ionosphere to \lya\ was negligible compared to that produced by the SXRs. 

Contrary to this, \cite{Milligan2020MXFlares} investigated the ionospheric response to an X-class flare through measurement of both VLF signal amplitude and atmospheric conductivity taken from magnetometer data. Comparing these to the respective \lya\ and SXR profiles from EUVS-E and the \textit{X-Ray Sensor} (XRS; \citealt{Hanser1996GOES_XRS}) they noted an impulsive geomagnetic disturbance in the Y-component of the magnetic field in the magnetometer data that was temporally correlated with the peak of the Ly$\rm \alpha$ irradiance and that peaked a few minutes before the SXR irradiance. This implied that the incident \lya\ produced enhanced current systems in the ionosphere, localised in the E-region through increased ionization of NO.

Unlike \ion{H}{1} \lya\ measurements, the \lya\ line of Helium II (\ion{He}{2}; 304\AA) has been comparatively well observed, with routine images taken by the \textit{Atmospheric Imaging Assembly} onboard SDO (SDO/AIA; \citealt{Boerner2012AIA}), the \textit{Extreme Ultraviolet Imager} onboard the \textit{Solar Terrestrial Relations Observatory - Ahead} (STEREO-A/EUVI; \citealt{Wuelser2004STEREO_A_EUVI}), the \textit{Extreme Ultraviolet Imaging Telescope} onboard the \textit{Solar and Heliospheric Observatory} (SOHO/EIT; \citealt{Domingo1995SOHO, Delaboudini1995SOHOEIT}), and others. \ion{H}{1} \lya\ and \ion{He}{2} \lya\ are both formed in similar regions in the lower solar atmosphere during quiescent conditions and both are therefore sensitive to changing dynamics in the solar atmosphere due to solar flares \citep{Simoes2016HeII,Brown2018LymanModelling}. Consequently, given their greater availability, \ion{He}{2} observations have been used as a proxy for \ion{H}{1} \lya\ (\citealt{Auchere2005LyaHeII}). \cite{Gordino2022LyaHeII} examined the relationship between \ion{He}{2} and \lya\ for a range of solar phenomena such as filament eruptions, prominences, and active regions, but did not include flare-related emission in their study. 

In this paper we present an multi-wavelength analysis of three solar flares of identical GOES class, but with significantly different \lya\ and \ion{He}{2} responses. HXR spectroscopy was used to determine whether differing nonthermal electron distributions are responsible for driving the different responses in EUV irradiance, and what fraction of the nonthermal energy is radiated by these two fundamental emission lines. A comparison was also made between the irradiance observations and semi-empirical model predictions from the \textit{Flare Irradiance Spectral Model} (FISM2; \citealt{Chamberlin2020FISM}). Section~\ref{sec:observations&analysis} outlines the multi-instrument observations taken for each flare in the sample and details the applied analysis. Section~\ref{sec:results} details the results of this analysis, and Section~\ref{sec:conclusion} summarises the findings from this study and discusses their broader implications.

\begin{figure*}[!ht]
\centering
\includegraphics[width=\textwidth]{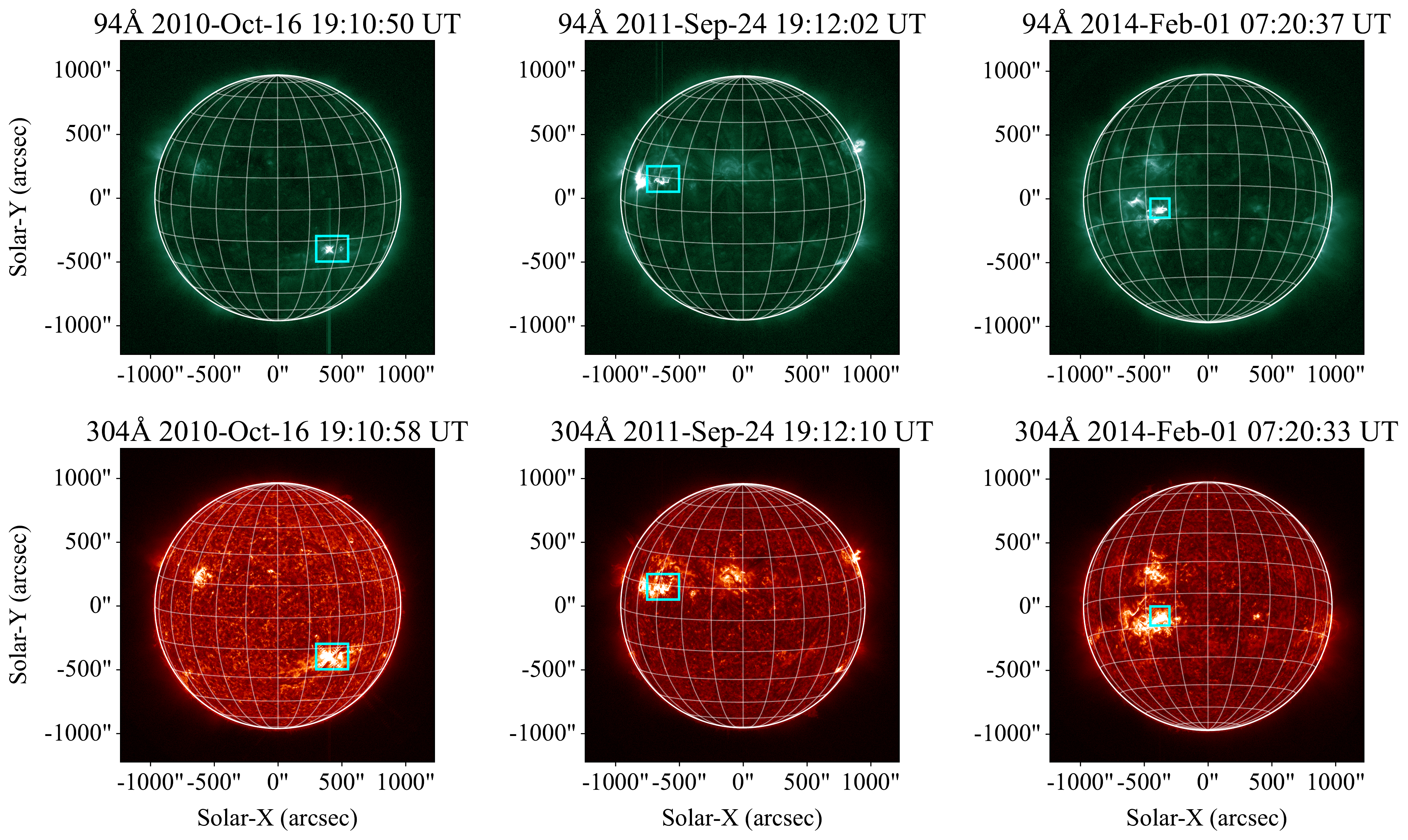}
\caption{Images taken from SDO/AIA for each flare listed in Table~\ref{tab:obs_summary} in the 94\AA\ (top row) and 304\AA\ (bottom row) filters approximately centred around their \lya\ peaks. Each column represents a single flare and the timestamp of each image is shown in the title above each frame. Cyan boxes in each image indicate the flaring region.}
\label{fig:aia_imaging}
\end{figure*}

\begin{table*}[!ht]
\begin{center}
\label{tab:obs_summary}
\caption{Summary of the observational characteristics of the flare sample.}   
\begin{tabular}{lccccc}
\hline
\hline
Solar Object Locator    &GOES Class	&Heliographic Position  &GOES start time    &GOES peak time &GOES end time \\
                        &           &                       &(UT)               &(UT)           &(UT)           \\
\hline
SOL2010-10-16T19:07:00  &M2.9	&S19W29		&19:07:00   	&19:12:30	    	&19:15:00           \\
SOL2011-09-24T19:09:00  &M3.0	&N13E45		&19:09:00	    &19:21:20   		&19:41:00           \\
SOL2014-02-01T07:14:00  &M3.0	&S14E17		&07:14:00 	    &07:23:47      		&07:36:00           \\
\hline
\end{tabular}
\end{center}
\end{table*} 

\begin{figure*}[ht]
\centering
\includegraphics[width=\textwidth]{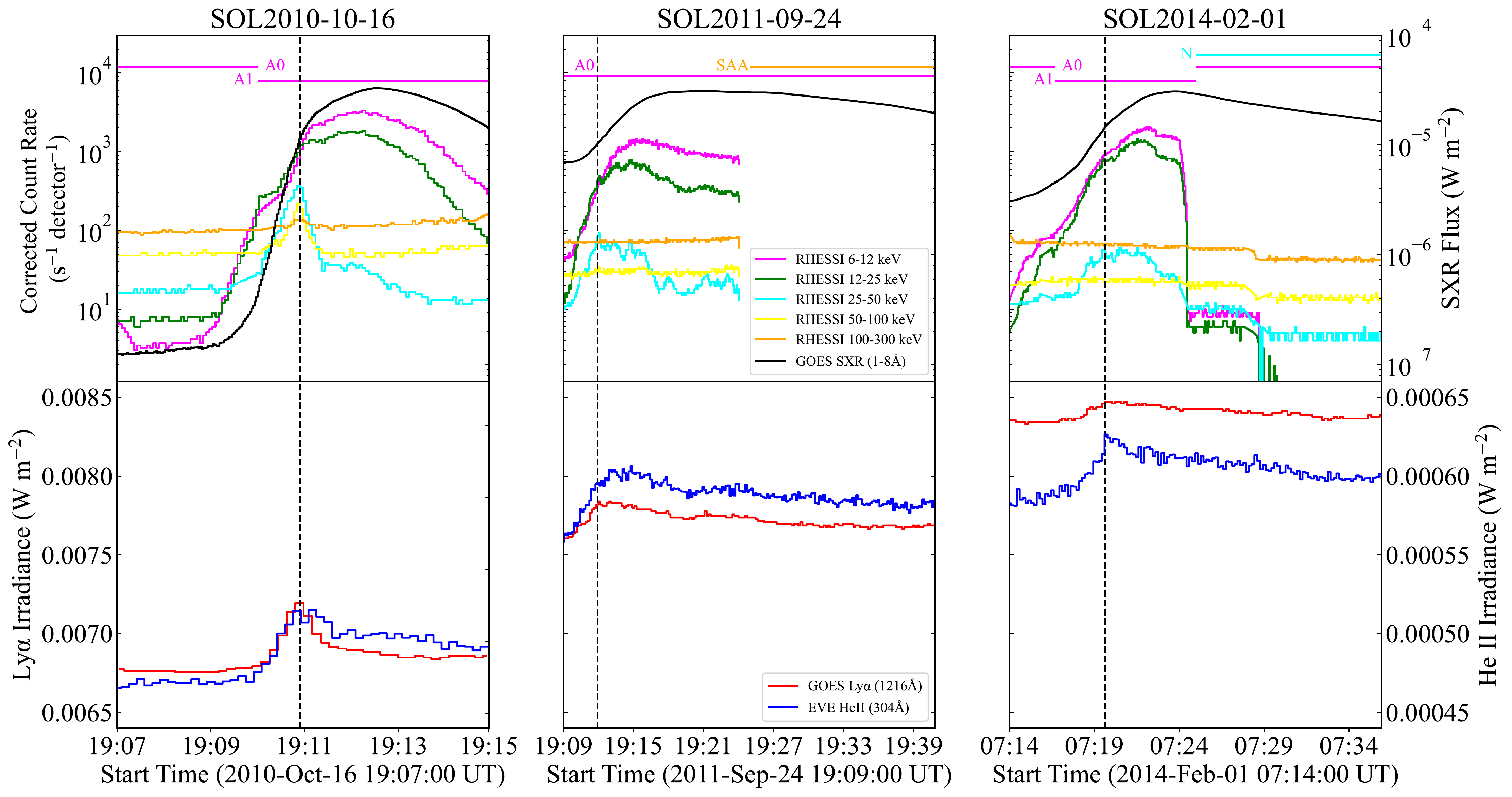}
\caption{Top Row: HXR lightcurves from RHESSI in discrete energy bands. The A0 and A1 flags (horizontal magenta lines) denote the RHESSI attenuator states. The A0 state corresponds to both thick and thin aluminum attenuators out of the detector field of view, the A1 state corresponds to thin in and thick out. The N flag (horizontal cyan line) denotes RHESSI night, where the spacecraft field of view is occulted by Earth. The SAA flag (horizontal orange line) denotes instrument switch off due to crossing of the South Atlantic Anomoly. SXR lightcurves from GOES/XRS in the 1--8\AA\ wavelength range (black solid). Bottom Row: Observed \lya\ (red) and \ion{He}{2} (blue) irradiance from GOES/EUVS-E and SDO/EVE, respectively. The vertical dashed line in each panel represents the peak of the 25--50~keV band for each flare.}
\label{fig:raw_lightcurves}
\end{figure*}


\section{Observation \& Analysis} \label{sec:observations&analysis}

This study focuses on the analysis of data from three $\rm \sim$M3 flares that occurred during Solar Cycle 24. Images of each flare in 94\AA\ and 304\AA\ emission taken by SDO/AIA are shown in the top and bottom rows of Figure~\ref{fig:aia_imaging}, respectively. The times and locations of each flare are summarized in Table~\ref{tab:obs_summary}. These flares were selected based on the following criteria:

\begin{itemize}

  \item The flares are of equivalent GOES class to remove any potential dependence of EUV irradiance variability on flare magnitude.

  \item Each flare occurred on-disk, reducing any opacity or foreshortening effects that may impact the measurement of optically-thick emission (center-to-limb variation, CLV; \citealt{Milligan2020MXFlares}).

  \item The impulsive phase of each flare must have been jointly observed by \textit{Reuven Ramaty High-Energy Solar Spectroscopic Imager} (RHESSI; \citealt{Lin2002RHESSI}), GOES-15/EUVS-E, and SDO/EVE MEGS-A to determine the nonthermal electron distribution assumed to be responsible for driving the associated increases in \lya\ and \ion{He}{2} irradiance.

  \item The flares must have occurred at times when the observations from GOES-15/EUVS-E were unaffected by geocoronal absorption, which may attenuate \lya\ emission (see \citealt{Meier1976Geocorona, Baluikin2019Geocorona}).
  
\end{itemize}

The top row of Figure~\ref{fig:raw_lightcurves} shows the SXR (black) and HXR (colored) lightcurves for each flare from GOES-15/XRS and RHESSI, respectively, while the bottom row shows the full-disk irradiance lightcurves in \lya\ (red curve) and \ion{He}{2} (blue curve), from GOES-15/EUVS-E and SDO/EVE, respectively across the full-flare duration. The vertical dashed line in each panel represents the peak of the 25--50~keV emission observed by RHESSI.

\subsection{GOES/XRS \& EUVS}\label{subsec:goes}

The GOES series of spacecraft operated by the \textit{National Oceanic and Atmospheric Administration} (NOAA) have offered continuous observations of solar emissions since the launch of GOES-1 in 1975. For this study, GOES-15 provided observations of SXR and \lya\ irradiance for each flare from the XRS and EUVS instruments, respectively. The XRS instrument consists of two channels observing in the 0.5--4\AA\ and 1--8\AA\ wavelength ranges, respectively, at 2s cadence. The peak flux of the 1-minute averaged SXR observations in the 1--8\AA\ wavelength range are the standard classification system for solar flare magnitudes (e.g. a peak flux of $3.0\times10^{-5}$~Wm$^{-2}$ corresponds to an M3.0 flare). As of 2020~October~28, GOES-8--15 SXR data accessed using the GOES object in the \textit{SolarSoftWare} system (SSWIDL; \citealt{Freeland1998SSWIDL}) are returned as ``true flux" values, meaning a scaling factor must be applied to these data to return the values used for determining flare class. A scaling factor of 0.7 has been applied to the SXR lightcurves presented in the top row of Figure~\ref{fig:raw_lightcurves} to preserve the GOES class of each flare.

The EUVS instrument is comprised of five channels (A--E) spanning the 50--170\AA, 240--340\AA, 200--620\AA, 200--800\AA, and 1180--1250\AA\ wavelength ranges. The E-channel (EUVS-E) is a dedicated channel centred around the \lya\ line at 1216\AA, taking broadband, full-disk irradiance measurements of the Sun at a cadence of 10.24s. The daily average values in \lya\ are scaled to those taken by SORCE/SOLSTICE to account for degradation in the EUVS-E instrument over time. The observed \lya\ emission for each flare is given by the red line in the bottom row of Figure~\ref{fig:raw_lightcurves}. The background \lya\ flux for each flare was determined by calculating the mean flux over a period of approximately 20 minutes prior to the flare onset. The data were then normalized to the background flux in order to quantify the enhancement in \lya\ irradiance attributed to the flare.

In order to calculate the energy radiated in \lya\ and SXRs, the irradiance data measured at Earth ($\rm I_{Earth}$) in units of W~m$^{-2}$ was converted to power radiated at the Sun (P$_{Sun}$) in units of erg~s$^{-1}$ using the expression: 

\begin{equation}
P_{Sun} \ = \ 2\pi R^{2}\times10^{7}~I_{Earth} \ \rm erg~s^{-1}
\label{eqn:wm2toergs}
\end{equation}

\noindent
where $R$ is the Sun-Earth distance corrected for deviations in Earth's orbit around the Sun at the time of each flare using the SSWIDL routine \texttt{get\_sun.pro}. The value $10^{7}$ is a conversion factor from J to erg. The total energy radiated during the impulsive phase was found by integrating the power from GOES start $\rightarrow$ GOES peak, allowing for a like-for-like comparison between the energy radiated in \lya\ and the energy deposited into the chrompshere by nonthermal electrons (see Section~\ref{subsec:rhessi}). This integration window was also extended to the full-flare period (GOES start $\rightarrow$ GOES end) to quantify the additional energy radiated after the initial injection of energy by the electrons, which may reveal evidence of further contributions to the overall emission in the latter stages of the flare. It should be noted that the GOES start time is defined as the first of four consecutive 1-minute intervals in which the 1--8\AA\ flux is monotonically increasing, with the flux in the fourth minute being 1.4 times the initial flux. The GOES end time is defined as the time at which the SXR flux reaches half of the peak value.\footnote{\url{https://www.ngdc.noaa.gov/stp/solar/solarflares.html}} The uncertainty in the total energy of the radiated components was calculated as the standard deviation in the flux of the background period.

\subsection{SDO/EVE}\label{subsec:sdo}

The EVE instrument onboard SDO acquires full-disk EUV spectra at a cadence of 10s through the A and B components comprising the \textit{Multiple EUV Grazing Spectrographs} (MEGS). The MEGS-A spectral range is 60--370\AA\ covering the 304\AA\ \ion{He}{2} line and MEGS-B spans 370--1060\AA. From the launch of SDO until 2014~May~26 MEGS-A had a near 100\% duty cycle, after which a power anomaly led to an instrumentation failure. The \ion{He}{2} 304\AA\ line was the strongest emission line measured by MEGS-A and is the brightest emission line in the solar irradiance spectrum shortward of 1000\AA\ \citep{Simoes2016HeII}. EVE data for each flare was acquired from the \textit{Laboratory for Atmospheric and Space Physics} (LASP\footnote{\url{https://lasp.colorado.edu/eve/data_access/}}) database. These data are available in hour-long Level 2 FITS files with emission line irradiances at 10~s cadence (EVL data). The 304\AA\ line irradiances are derived from Level 2 spectral data integrated over the line centre ($\rm\Delta$\AA=2.5). These data were background subtracted, normalized, and converted using Equation~\ref{eqn:wm2toergs} using the same process as the GOES/EUVS-E data. The observed \ion{He}{2} emission for each flare is given by the blue line in the bottom row of Figure~\ref{fig:raw_lightcurves}.

\begin{figure*}[ht!]
\centering
\includegraphics[width=\textwidth]{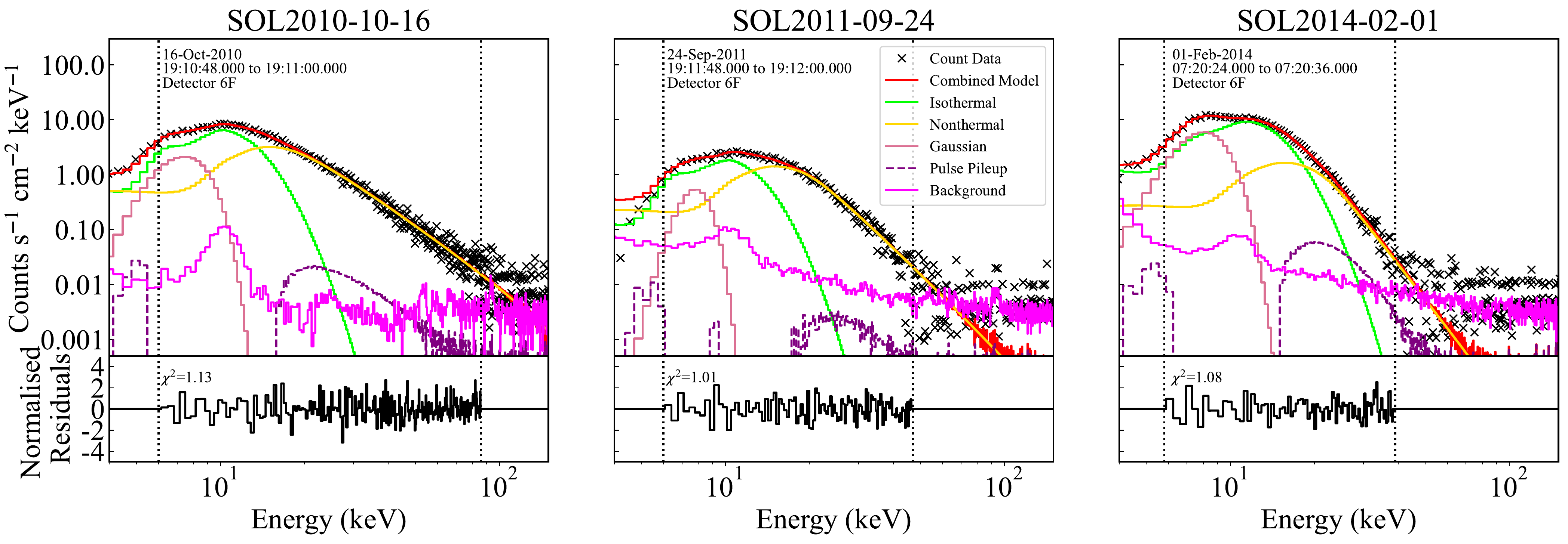}
\caption{RHESSI spectra from the HXR peak of each flare with individual fit components overlaid. Each plot shows the model fits to the count data at the peak of the impulsive phase, taken from detector 6. The normalized residuals are shown in the lower panels. The dotted lines indicate the energy range over which the fits were performed.}
\label{fig:rhessi_fit_examples}
\end{figure*}

\subsection{FISM2}\label{subsec:FISM2}

Prior to the launch of SDO/EVE, synthetic EUV spectra could be generated using FISM \citep{Chamberlin2007FISM, Chamberlin2008FISMoriginal}, which is a semi-empirical model of solar UV emission in the 0--1900\AA\ wavelength range, designed to fill temporal and spectral gaps in observations of solar flare irradiance. These proxy spectra could serve as inputs into planetary atmospheric models such as \textit{Whole Atmosphere Community Climate Model with thermosphere and ionosphere eXtension} (WACCM-X; \citealt{Garcia2007WACCMX, Marsh2013WACCM, Neale2013WACCMX, Qian2018WACCMX, Solomon2018WACCMX}), which models fluctuations in electron density, vertical ion drift, zonal electric field, and conductance in the ionosphere due to solar activity. Following the launch of SDO, irradiance profiles generated by FISM2 are based on data acquired by SDO/EVE and by the\textit{X-ray Photometer System} (XPS) and SOLSTICE instruments onboard SORCE. The data from FISM2 are provided in 1\AA\ spectral bins and flare products are produced with a 60s cadence. For this study, FISM2 was employed to provide predicted irradiance profiles in both \lya\ and \ion{He}{2} for each flare, which could then be directly compared to the measured profiles from GOES/EUVS-E and SDO/EVE. Irradiance profiles were acquired for each of the three flares from the \textit{LASP Interactive Solar Irradiance Datacentre} (LISRD\footnote{\url{https://lasp.colorado.edu/lisird/}}). FISM2 data from 1180--1250\AA\ was scaled to the instrument response data for EUVS-E and integrated to mimic the broadband \lya\ observations from GOES-15. For the \ion{He}{2} profiles, FISM2 data was summed over the bins surrounding the line centre to resemble the resolution of the irradiance data from SDO/EVE. These were then background subtracted and normalized using identical processes and periods to their respective observations.

\subsection{RHESSI}\label{subsec:rhessi}

Nonthermal electrons incident on the chromosphere during solar flares drive the emission of HXR bremsstrahlung through interaction with the dense plasma at the flare footpoints \citep{Dennis&Schwartz1982FlareBible}. RHESSI provides high-resolution imaging and spectroscopic measurements of HXR emission in flares that can be used to discern the properties of the nonthermal electrons assumed to be responsible for driving impulsive emission at longer wavelengths. HXR lightcurves for the three flares in this study are given by the colored lines in the top row of Figure~\ref{fig:raw_lightcurves}, where each line represents each of the discrete energy bands ranging from 6--300~keV. Assuming the HXR emissions are driven by thick-target interactions, the electron energy spectrum can be derived from the count spectra under the assumptions of the Collisional Thick-Target Model (CTTM; \citealt{Brown1971CTTM, Brown1972CTTM, Brown1973CTTM}). 

The total power in the electron distribution is given by:
\begin{equation}
P_{nth}(E \geq E_{c}) \ = \ \int\limits_{E_{c}}^\infty E F(E) \ dE  \ \rm  erg \ s^{-1}  
\label{eq:pnth_integral}
\end{equation}

\noindent
where $E_{c}$ is the low energy cutoff and the electron energy flux distribution, F(E), can be written in power-law form $AE^{-\delta}$. The coefficient, $\rm A$, corresponds to the electron number flux propagating toward the chromosphere and $\rm \delta$ represents the spectral index of the distribution. The measured value of $\rm E_{c}$ is taken as an upper limit to the nonthermal cutoff due to the dominance of thermal emission at low energies. Subsequently, nonthermal electron energy calculated through the HXR spectral fitting is considered to be a lower limit to the total energy (\citealt{Holman2003ElectronBremmHXRSpectra, Ireland2013ModellingHXRFlares}). With these, Equation~\ref{eq:pnth_integral} then becomes:
\begin{equation}
P_{nth}(E \geq E_{c}) \ = \ \frac{\kappa_{E}A}{(\delta - 2)} E_{c}^{(2-\delta)} \ \rm  erg \ s^{-1} 
\label{eq:pnth_post_integration}
\end{equation}

\noindent
where $\rm \kappa{_{E}}$ is a conversion factor from keV to erg. For each flare, RHESSI count spectra were compiled for seven detectors (omitting detectors 2 and 7 due to their comparatively reduced sensitivity) using native energy binning from 3--300~keV.

\begin{figure*}[!ht]
\centering
\includegraphics[width=\textwidth]{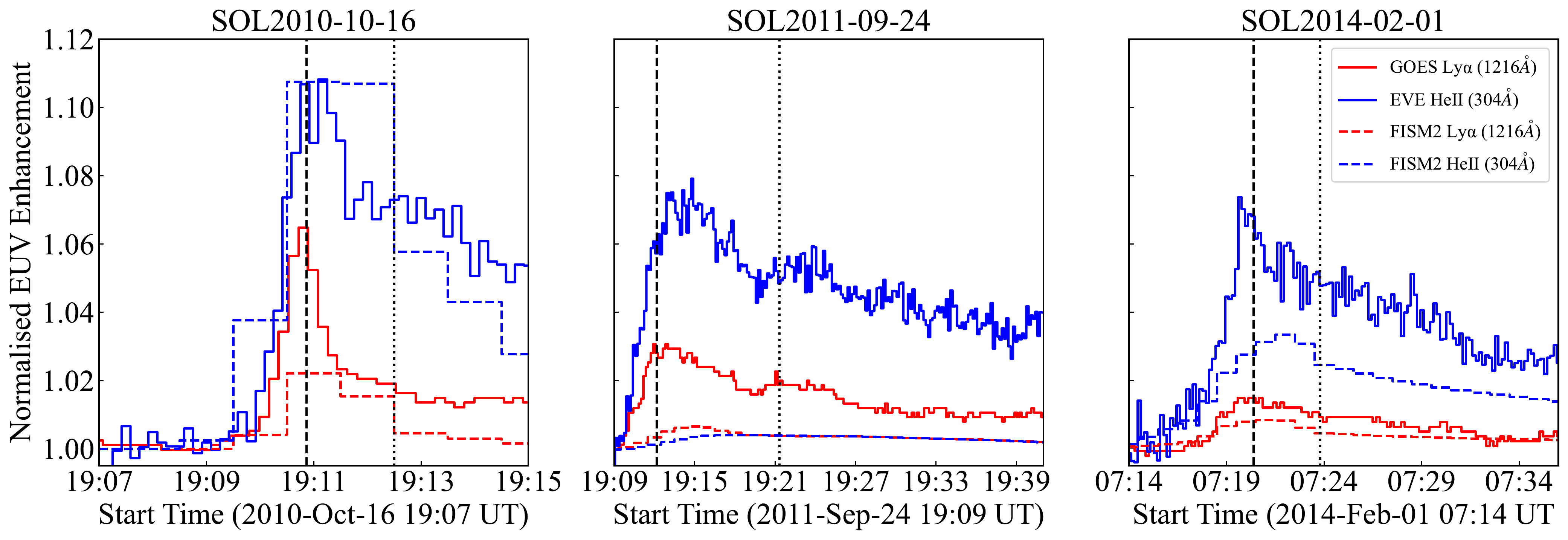}
\caption{Normalized irradiance profiles for each of the three flares. Red and blue curves correspond to \lya\ and \ion{He}{2} data, respectively. Solid lines correspond to observational data from GOES/EUVS-E and SDO/EVE while dashed lines correspond to the predicted irradiance profiles from FISM2. Overplotted are vertical lines corresponding to the peaks of the 25--50~keV emission from RHESSI (dashed black) and the 1--8\AA\ emission from GOES/XRS (dotted black).}
\label{fig:irradiance_obs_FISM}
\end{figure*}


\begin{deluxetable*}{ccccccc}[!ht]
\tablecolumns{7}
\tablewidth{0pt}
\tablecaption{Peak irradiance enhancements in \lya\ and \ion{He}{2} for each flare, both as a percentage normalized to the background flux and as an absolute background-subtracted value.    \label{tab:irradiances}}
\tablehead{\colhead{} & \multicolumn{2}{c}{Peak \lya} & \multicolumn{2}{c}{Peak \ion{He}{2}} & \colhead{Peak \lya} & \colhead{Peak \ion{He}{2}}\\
\vspace{-0.7cm}\\
\colhead{Flare} & \multicolumn{2}{c}{Enhancement} & \multicolumn{2}{c}{Enhancement} & \colhead{Irradiance ($\rm W$~$\rm m^{-2})$} & \colhead{Irradiance ($\rm W$~$\rm m^{-2})$} \\
\cline{2-7}
\colhead{} & \colhead{GOES/EUVS-E} & \colhead{FISM2} & \colhead{SDO/EVE} & \colhead{FISM2} & \colhead{GOES/EUVS-E} & \colhead{SDO/EVE}}
\startdata
\vspace{-0.2cm} \FLone & $6.4\%$ & $2.2\%$ & $10.8\%$ & $10.7\%$ & $4.4\times10^{-4}$ & $5.1\times10^{-5}$ \\
\vspace{-0.2cm} \\
\vspace{-0.2cm} \FLtwo & $3.3\%$ & $0.7\%$ & $7.4\%$ & $0.4\%$ & $2.4\times10^{-4}$ & $4.3\times10^{-5}$ \\
\vspace{-0.2cm} \\
\vspace{-0.2cm} \FLthree & $1.5\%$ & $0.8\%$ & $7.3\%$ & $3.3\%$ & $1.3\times10^{-4}$ & $4.3\times10^{-5}$ \\
\vspace{-0.2cm} \\
\enddata
\end{deluxetable*}

The count spectra for each flare were binned into 12s intervals to match the EUVS-E cadence to the nearest integer number of RHESSI rotations. Each spectrum was then fit with a combination of an isothermal component, a Gaussian, and a nonthermal CTTM electron spectrum. An additional component was also included to account for pulse pile-up using the most up-to-date version available in OSPEX. The spectra were also corrected for
albedo attributed to Compton backscattered photons within the OSPEX software (\citealt{Kontar2006AlbedoBackScatter}). Spectra from the HXR peak of each flare are shown in Figure~\ref{fig:rhessi_fit_examples} along with their fit components. For each interval, the power in the nonthermal electrons was calculated from Equation~\ref{eq:pnth_post_integration}, where the low energy cutoff, spectral index, and electron rate at each time interval are defined by the nonthermal component fitted to the spectra. These were then integrated in time to give the total nonthermal energy in each flare. The nonthermal power over the flare duration was calculated for each detector from the parameters at each interval. This was then integrated to return the total nonthermal energy for each detector. The total nonthermal energy in each flare was taken as the detector-averaged energy $\rm\pm$ the standard deviation across the detectors.

\section{Results} \label{sec:results}

\subsection{Irradiance Variability}\label{subsec:irradiance_variability}

The full-flare irradiance profiles for \lya\ (red) and \ion{He}{2} (blue), normalized to their respective backgrounds, are shown in Figure~\ref{fig:irradiance_obs_FISM}. Solid curves denote the observed profiles while dashed lines denote the profiles produced by FISM2. From the GOES-15/EUVS-E data, \lya\ enhancements of 6.4\%, 3.3\%, and 1.5\% were found for \FLone, \FLtwo, and \FLthree, respectively. The corresponding \ion{He}{2} enhancements from SDO/EVE appear significantly larger than the \lya\ across all three flares with values of 10.8\%, 7.4\%, and 7.3\%, respectively. This is likely due to the much higher background flux in \lya. \FLone\ was found to have both the largest \lya\ and \ion{He}{2} enhancements. \FLtwo\ and \FLthree\ had significantly different \lya\ enhancements yet their \ion{He}{2} enhancements were remarkably similar, differing by only 0.1\%. For all three flares, the peaks in their \lya\ emission coincided with the peaks of the HXR emission as measured by RHESSI (dashed vertical lines in Figure~\ref{fig:irradiance_obs_FISM}), suggesting that these impulsive enhancements were driven by nonthermal electron incidence. This is in agreement with a number of previous studies that report a temporally correlated relationship between \lya\ and nonthermal emission (\citealt{Nusinov2006LyaNTCorr, deCosta2009NTLyaCorr, Milligan2017NTLyaCorr, Dominique2018NTLyaCorr, Lu2021LymanNTCorr, Li2022LyaNTCorr}). However, each flare also exhibits a prolonged decay time before returning to background levels. The peak \lya\ and \ion{He}{2} enhancements for each flare are summarised in Table~\ref{tab:irradiances}.

Figure~\ref{fig:irradiance_obs_FISM} also shows that the peak flare enhancements predicted by FISM2 (dashed red and blue lines) are consistently unrepresentative of the observed values. While FISM2 also predicts some variation in peak \lya\ enhancement between flares, these enhancements are $<$2.5\% in all cases. The range of \ion{He}{2} peak values from FISM2 is much broader than the observations. A good estimate was made for \FLone, however the irradiances of \FLtwo\ and \FLthree\ were significantly under-predicted. The values derived from FISM2 are also presented in Table~\ref{tab:irradiances}.

Several factors may contribute to the disagreement between FISM2 and observational data. For one, FISM2 flare models rely on the time derivative of GOES/XRS observations as a proxy for the impulsive phase of each flare. This means that the impulsivity of each flare will influence the FISM2 estimations of the associated emission. \FLone\ clearly demonstrates a more impulsive profile compared to \FLtwo\ and \FLthree, therefore while their SXR magnitudes are equivalent, the magnitude of their SXR derivatives is varied. Differing associated background levels further drive variation in the SXR derivative magnitude. Given the flares presented here are of an intermediate class, their associated background level will have a more significant impact on the calculated SXR derivative compared to larger (X-class) flares. Moreover, the observing statistics used in the FISM2 data sets are biased towards smaller flares due to their higher occurrence rates. Consequently, the lack of data for larger flares with good signal-to-noise measurements leads to lower estimations in the associated flux.

When using SXRs as a proxy for optically-thick emission, a correction must be applied to account for CLV effects dependent on flare location. The method for correcting for CLV in FISM2 is consistent with that of the first version of FISM with an increased data set. The full set of algorithms and function parameters used for computing a CLV correction, as well as a detailed presentation of the method, are presented in \cite{Chamberlin2008FISMoriginal}. As FISM2 is an empirical model, its accuracy is limited by the observations on which it depends. The high resolution spectral measurements ($\rm \lambda<0.1~nm$) used for FISM2 are taken from the SORCE/SOLSTICE scanning spectrograph, which has missed a large number of flares. Consequently, while there are a statistically sufficient number of observations of the gradual phase from SORCE/SOLSTICE to compute a CLV correction in the 115--190~nm range, a lack of observations means the same could not be done for the impulsive phase in this range. Ultimately, this will impact the \lya\ irradiance profiles from FISM2. However, given the constraints on flare position discussed in Section~\ref{sec:observations&analysis}, the flares in this study should experience minimal CLV effects due to their positions on-disk (see \citealt{Milligan2020MXFlares}). In spite of this, the \ion{He}{2} irradiance profiles from FISM2, which \textit{have} been treated with a CLV correction, still demonstrate underestimation of the observed emission for two flares in the sample. This suggests further that the fundamental processes involved in producing the FISM2 irradiance profiles must be improved upon for future iterations of FISM.

\begin{figure*}[!ht]
\centering
\includegraphics[width=\textwidth]{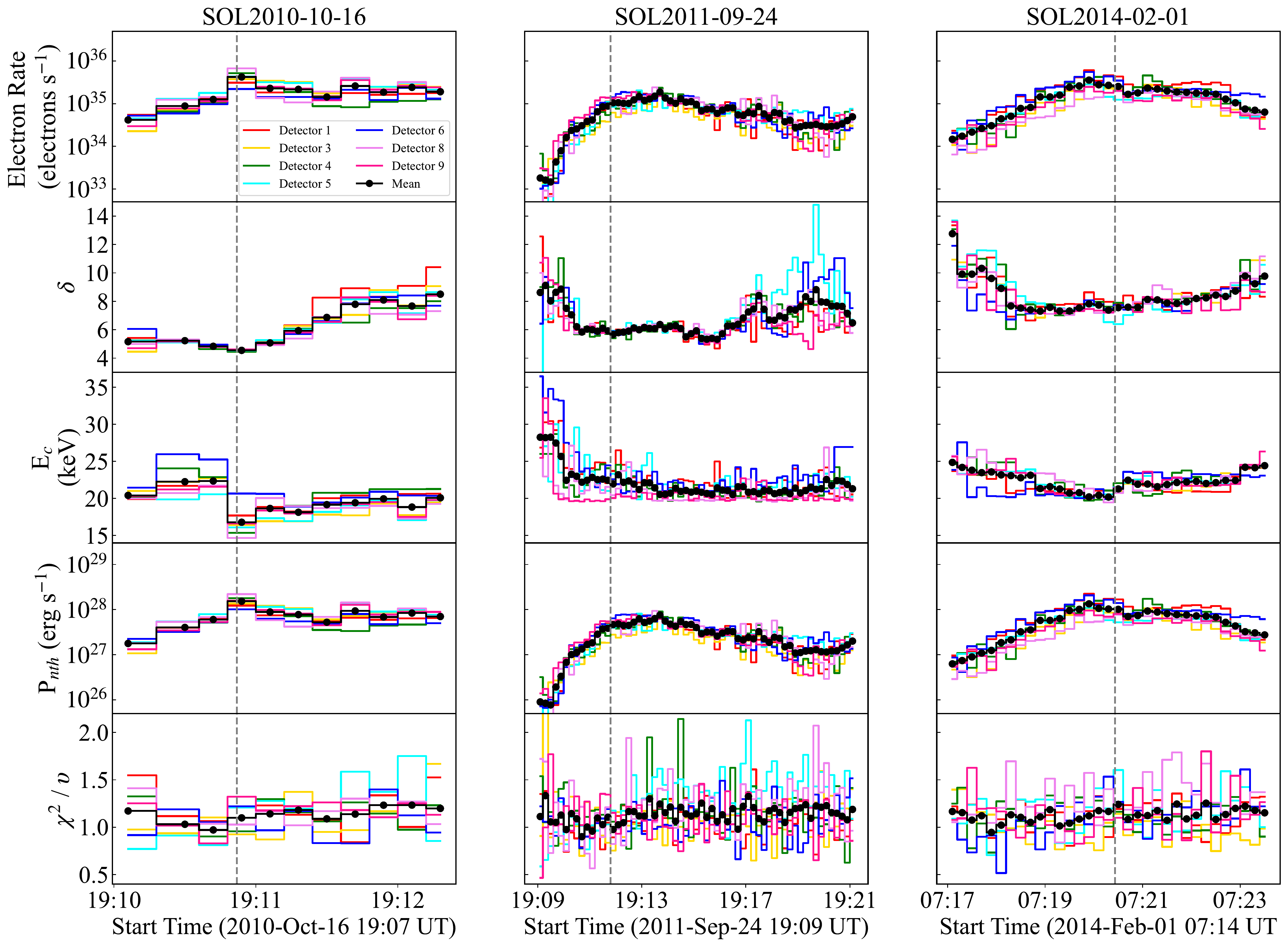}
\caption{Temporal evolution of the nonthermal parameters for each flare as derived from RHESSI HXR spectroscopy. First row: electron rate ($\rm A$). Second row: spectral index ($\rm \delta$). Third row: low energy cutoff ($\rm E_{c}$). Fourth row: nonthermal electron power ($\rm P_{nth}$). Fifth row: reduced $\rm \chi^{2}$ value for the combined fit model at each interval. The vertical dashed grey line indicates the peak time of the RHESSI 25--50~keV energy band.}
\label{fig:nonthermal_fit_parameters}
\end{figure*}

\begin{deluxetable*}{ccccc}[!ht]
\tablecolumns{4}
\tablewidth{0pt}
\tablecaption{Values of nonthermal fit parameters at the peak of the 25--50~keV HXR emission and the associated peak nonthermal electron power.\label{tab:nonthermal_parameters}}
\tablehead{
\colhead{Flare}  & \colhead{$\rm A$} & \colhead{$\rm \delta$} & \colhead{$\rm E_{c}$} & \colhead{$\rm P_{nth}$}\\
\vspace{-0.7cm} \\
\colhead{} &\colhead{($\rm 10^{35}$~electrons$\rm \ s^{-1}$)} & \colhead{} & \colhead{($\rm keV$)} & \colhead{$(\rm 10^{28}$~$\rm erg$~$\rm s^{-1})$}}
\startdata
\vspace{-0.2cm} \FLone  & $4.2\pm1.3$ & $4.6\pm0.1$ & $16.8\pm1.8$ & $1.5\pm0.4$ \\
\vspace{-0.2cm} \\
\vspace{-0.2cm} \FLtwo & $1.1\pm0.3$ & $5.7\pm0.2$ & $21.8\pm1.3$ & $0.5\pm0.1$  \\
\vspace{-0.2cm} \\
\vspace{-0.2cm} \FLthree & $2.7\pm1.2$ & $7.4\pm0.4$ & $20.2\pm0.5$ & $1.0\pm0.4$  \\
\vspace{-0.2cm} \\
\enddata
\end{deluxetable*}

\subsection{Nonthermal Electron Parameters}

From the fits to the HXR spectra from RHESSI, the parameters that describe the distribution of nonthermal electrons believed to be responsible for driving the impulsive increases in EUV irradiance were derived. Figure~\ref{fig:nonthermal_fit_parameters} shows the progression of $\rm A$, $\rm \delta$, $\rm E_{c}$, and nonthermal electron power ($\rm P_{nth}$) across the impulsive phase for each flare. The reduced $\chi^{2}$ values are shown in the bottom row. The value of each parameter was examined at the peak of the 25--50~keV energy band (and therefore the peak of the \lya\ emission), indicated by the vertical grey dashed lines in Figure~\ref{fig:nonthermal_fit_parameters}. These are summarised in Table~\ref{tab:nonthermal_parameters} along with the corresponding $\rm P_{nth}$ value.

An apparent soft-hard-soft (SHS) behavior is observable from the temporal evolution of $\rm \delta$ across \FLone\ (left panel in row 2 of Figure~\ref{fig:nonthermal_fit_parameters}), where the value of $\rm \delta$ decreases (the spectrum hardens) as the flare reaches its peak HXR flux, before increasing (the spectrum softens) once more after the HXR peak time. \FLtwo\ shows similar but less pronounced SHS behavior compared to \FLone\ (middle panel in row 2 of Figure~\ref{fig:nonthermal_fit_parameters}) . In \FLthree, this behavior is particularly weak (right panel in row 2 of Figure~\ref{fig:nonthermal_fit_parameters}). This temporal behavior may give indication to the nature of chromospheric heating in each flare driven by nonthermal electrons. The SHS behavior is an intrinsic property of solar flares characterized by the tendency of flare-related HXR emission to vary with spectral hardness (\citealt{GrigisBenze2004SHS}). A prominent SHS component with a short temporal duration implies a more impulsive acceleration event, such as in the case of \FLone, whereas a longer duration and less pronounced SHS behavior implies a more gradual nature, such as in the case of \FLtwo\ and \FLthree. 

\begin{figure*}[!ht]
\centering
\includegraphics[width=\textwidth]{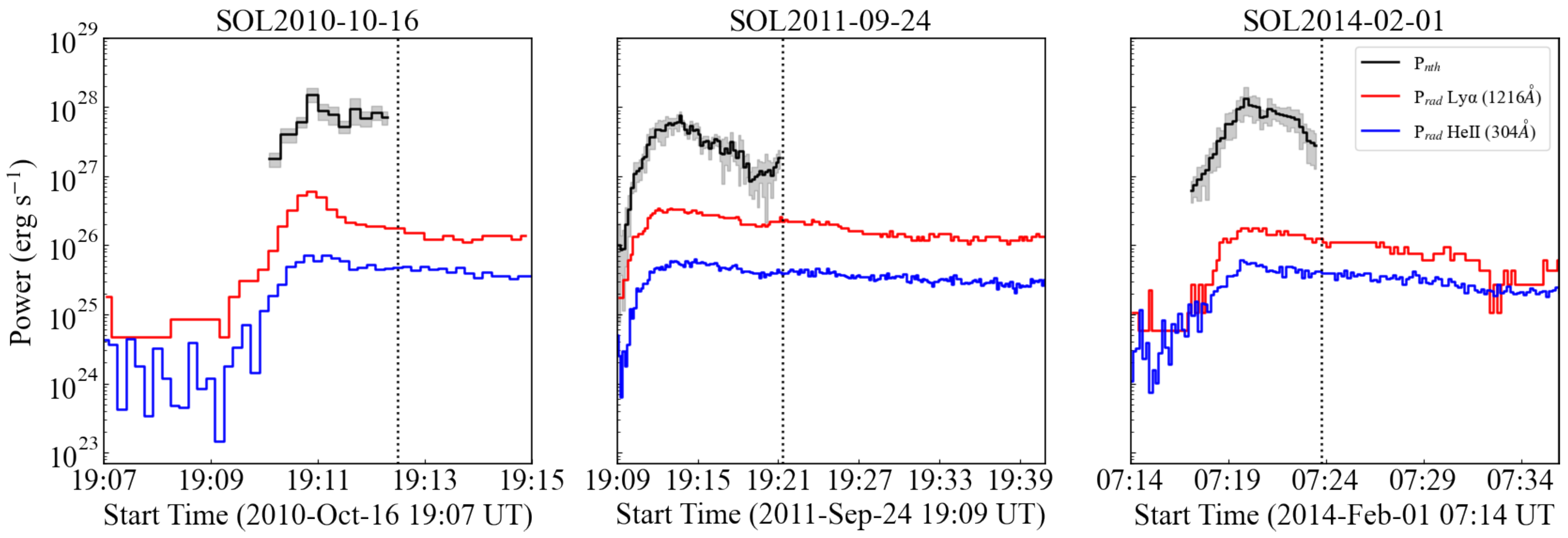}
\caption{Nonthermal and radiated power as a function of time for each flare in the sample. The dashed lines indicate the impulsive phase integration limits. The shaded region represents $\rm P_{nth} \pm \sigma$, where $\rm \sigma$ is taken as the as the standard deviation of $\rm P_{nth}$ over the full set of RHESSI detectors used.}
\label{fig:energetics}
\end{figure*}

\begin{deluxetable*}{ccccccc}[!ht]
\tablecaption{Total energy in nonthermal electrons and radiated components during each flare. \label{tab:energies}}
\tablehead{\colhead{Flare} & \colhead{Time Range (UT)} & \multicolumn{3}{c}{Total Energy (erg)} & \multicolumn{2}{c}{$\rm \sfrac{E_{rad}}{E_{nth}}$ (\%) }\\
\cline{3-7}
\colhead{} & \colhead{} & \colhead{$\rm E_{nth}$} & \colhead{$\rm E_{Ly\alpha}$} & \colhead{$\rm E_{He~II}$} & 
\colhead{Ly$\rm \alpha$} & \colhead{\ion{He}{2}}}
\startdata
\vspace{-0.2cm} \FLone & 19:07:00~-~19:12:30 & $9.7\pm1.6\times10^{29}$ & $4.3\times10^{28}$ & $7.1\times10^{27}$ & $4.4\pm0.7$ & $0.7\pm0.1$ \\
\vspace{-0.2cm} \\
\vspace{-0.2cm}  & 19:07:00~-~19:15:00 &  & $6.3\times10^{28}$ & $1.3\times10^{28}$  &  &  \\
\vspace{-0.2cm} \\
\cline{1-7}
\vspace{-0.2cm} \FLtwo & 19:09:00~-~19:21:20 & $2.0\pm0.3\times10^{30}$ & $1.6\times10^{29}$ & $2.9\times10^{28}$  & $7.9\pm1.1$ & $1.4\pm0.2$ \\
\vspace{-0.2cm} \\
\vspace{-0.2cm}  & 19:09:00~-~19:41:00 &  & $3.3\times10^{29}$ & $6.6\times10^{28}$  &  &  \\
\vspace{-0.2cm} \\
\cline{1-7}
\vspace{-0.2cm} \FLthree & 07:14:00~-~07:23:47 & $2.3\pm0.7\times10^{30}$ & $4.7\times10^{28}$ & $1.5\times10^{28}$  & $2.0\pm0.6$ & $0.6\pm0.2$ \\
\vspace{-0.2cm} \\
\vspace{-0.2cm} & 07:14:00~-~07:36:00 &  & $9.7\times10^{28}$ & $3.6\times10^{28}$  & & \\
\vspace{-0.2cm}
\enddata
\end{deluxetable*}


Considering the findings from Section~\ref{subsec:irradiance_variability}, there is an apparent tendency for the flux enhancement in \lya\ to scale with the value of $\rm \delta$, such that a smaller $\rm \delta$, and therefore harder electron energy spectrum, correlates to a larger enhancement in \lya\ irradiance. \FLone\ had a $\rm \delta$ value of 4.6 at the peak of the HXR emission and showed a 6.4\% enhancement in \lya. Whereas \FLthree\ had a $\rm \delta$ value of 7.4 at the peak of the respective 25--50~keV HXRs and showed a 1.5\% \lya\ enhancement.

The value of A at the peak of the 25--50~keV HXRs does not seem to have any relationship to the enhancements in EUV emission. Similarly, the value of $\rm E_{c}$ at the 25--50~keV peak does not appear to show a link to the enhancement in \lya\ emission. An $\rm E_{c}$ of 16.8, 21.8, and 20.2~keV was found found for \FLone, \FLtwo, and \FLthree, respectively. Here we see that the lowest value of $\rm E_{c}$ is attributed to the flare with the greatest enhancement in \lya, however the largest $\rm E_{c}$ value is found in the flare with the second largest enhancement in \lya. It is noted that there is a particular ambiguity in determining the value of $\rm E_{c}$, especially for weaker flares, due to the dominance of thermal emission at low energies. Thus, the apparent relationship between $\rm \delta$ and the \lya\ flux is the most compelling result here. 

\lya\ emission is optically-thick and may form over a range of chromospheric altitudes. Using RADYN (\citealt{Allred2005RADYN, Allred_2015}) simulations of the chromospheric response to flare heating, \cite{Brown2018RADYN_Lyman} found that variation in electron beam parameters may impact plasma upflows in the chromosphere and subsequently may effect the emission and absorption of \lya\ and other lines in the Lyman series. Thus for the flares studied here, the variability in nonthermal parameters found from HXR spectral analysis may directly influence the chromospheric dynamics in each flare, in turn impacting the levels of associated \lya\ emission. 

Comparing Table~\ref{tab:nonthermal_parameters} to Table~\ref{tab:irradiances}, it is apparent that the value of $\rm P_{nth}$ has no clear relationship with the peak enhancement in either \lya\ or \ion{He}{2}. While \FLone\ shows both the largest peak $\rm P_{nth}$ and the largest peak in \lya\ and \ion{He}{2}, these values do not appear to be related in \FLtwo\ and \FLthree. This implies that the total energy deposited into the chromosphere does not necessarily dictate the degree to which \lya\ or \ion{He}{2} emission is enhanced. Instead, it is more likely that the depth at which the energy is deposited in the chromosphere will play a greater role in dictating the enhancement in the EUV emission, which will depend on the nonthermal properties themselves. 

\subsection{Energetics}

The power in nonthermal electrons derived from RHESSI fits for each flare is given by the black line in Figure~\ref{fig:energetics}, the red and blue lines show the power radiated in \lya\ and \ion{He}{2} calculated from  GOES/EUVS-E and SDO/EVE lightcurves, respectively. Here, the plot limits encapsulate the full-flare period; the vertical dotted line denotes the GOES peak and therefore the upper limit of the impulsive phase.

For \FLone, the total energy contained in nonthermal electrons  was found to be  $9.7\times10^{29}$~erg. The corresponding energy radiated in \lya\ during the impulsive phase (330s) was found to be $4.3\times10^{28}$~erg, accounting for 4.4\% of the total nonthermal energy budget. Extending the integration window to the entire GOES flare period, the total energy radiated in \lya\ was found to be $6.3\times10^{28}$~erg. The radiated energy in \ion{He}{2} was found to be $7.1\times10^{27}$~erg over the impulsive phase, amounting to 0.7\% of the nonthermal energy budget, and was found to increase to $1.3\times10^{28}$~erg over the full-flare period. Taking the ratio of the impulsive phase and full-flare period energies ($\rm\sfrac{E_{FP}}{E_{IP}}$), the radiated energies in \lya\ and \ion{He}{2} are found to increase by a factor of 1.5 and 1.8 between the impulsive phase and the entire flare period, respectively.

For \FLtwo, the impulsive phase lasted around 740s and the total nonthermal energy was found to be $2.0\times10^{30}$~erg. The energy radiated in \lya\ over the impulsive phase was found to be $1.6\times10^{29}$~erg accounting for 7.9\% of the total nonthermal energy. Integrating over the entire flare duration, the energy in \lya\ increased by a factor of 2.1 to $3.3\times10^{29}$~erg. For \ion{He}{2}, the radiated impulsive phase energy was $2.9\times10^{28}$~erg, amounting to 1.4\% of the nonthermal energy. Over the entire flare duration, the \ion{He}{2} energy was found to increase to $6.6\times10^{28}$~erg, which is 2.3 times the impulsive phase energy. 

Finally, for \FLthree\ the total nonthermal energy was found to be $2.3\times10^{30}$~erg over the 587s impulsive phase. The corresponding \lya\ energy radiated in this period was $4.7\times10^{28}$~erg accounting for 2.0\% of the nonthermal electron energy. The energy radiated over the entire flare period in \lya\ was found to increase to $9.7\times10^{28}$~erg. Thus, the value of $\rm\sfrac{E_{FP}}{E_{IP}}$ for \lya\ in \FLthree\ was 2.1. Once again, examining the \ion{He}{2} emission in the impulsive phase, the total energy radiated at this wavelength was $1.5\times10^{28}$~erg, 0.6\% of the nonthermal energy, increasing by a factor of 2.4 to $3.6\times10^{28}$~erg over the full-flare period.

 A full summary of the flare energetics is presented in Table~\ref{tab:energies}. The typical nonthermal energy was found to be approximately on the order of $10^{30}$~erg in reasonable agreement with previous studies of flare energetics in M-class flares \citep{stHilaire2005FlareEnergy}. It is clear from Table~\ref{tab:energies} that Ly$\rm \alpha$ may carry a reasonable portion of the energy deposited in the chromosphere by nonthermal electrons during the impulsive phase. \lya\ was found to radiate approximately 2--8\% of the energy deposited in the chromosphere by nonthermal electrons. The lower limit to this range is lower than value found by \cite{Milligan2014EUVEnergy} and the overall range is broader, this may be attributed to their use of the MEGS-P instrument within SDO/EVE to measure the \lya\ flux, which has been found to produce distinctly different \lya\ profiles compared to GOES/EUVS-E (\citealt{Milligan2016SDO_EVE_MEGSP}), as well as the differing magnitude of the flare examined. 
 
 Each flare showed elevated \lya\ and \ion{He}{2} flux during their decay phase. A value of $\rm\sfrac{E_{FP}}{E_{IP}}$~$\geq2$ implies at least half of the total energy radiated during the flare is radiated after the impulsive phase, such was the case for \FLtwo\ and \FLthree. Possible contributions to the elevated flux may be attributed to evaporated material in the corona cooling to chromospheric temperatures \citep{Jing2020CoolingLyaEmission}, failed filament eruptions \citep{Wauters2022M67Lya}, or evaporated material driven by thermal conduction \citep{Zarro1988CondEvap}, for example. While we cannot necessarily determine the source of the emission using disk-integrated measurements alone, non-chromospheric contributions should not be neglected from the interpretation of the results presented here. Ultimately these may contribute to total incident flux in the Earth's atmosphere attributed to each flare, as well as reveal information on the distribution of energy in the solar atmosphere during flares.

 
\section{Summary \& Conclusions} \label{sec:conclusion}

In this study, a multi-wavelength analysis was conducted for three solar flares with equivalent GOES magnitude but distinctly different \lya\ profiles. The observed \lya\ and \ion{He}{2} emission in each flare was compared to synthetic line profiles generated by FISM2, which despite some agreement for one flare, generally under-predicted the \lya\ and \ion{He}{2} flux. It was hypothesised that the variability in the associated \lya\ emission for each flare may be attributed to differing properties of the nonthermal electrons incident on the chromosphere. This was examined by carrying out spectral analysis of flare-related HXRs to discern a set of parameters that describe the energy distribution of the nonthermal electrons in each flare. Following this, the radiated energy in \lya\ and \ion{He}{2} was calculated for each flare and compared to the total nonthermal energy deposited into the chromosphere over the impulsive phase, determining the contribution of both wavelengths to the radiated energy budget.

The peak enhancement in \lya\ emission between the flare sample was found to range from 1.5--6.4\%. Comparatively, \cite{Milligan2021B&CClassFlares} reported an average enhancement in \lya\ emission of 1.5\% for M-class flares, placing two of the three flares in this study above that average. Interestingly, FISM2 estimated \lya\ enhancements of $<$2.5\% for each flare. 

The enhancement in \lya\ emission demonstrated a clear tendency to scale with the spectral index, with the largest \lya\ enhancement corresponding to the smallest spectral index value at the peak of the 25--50~keV HXR emission. No such behavior was found for the electron rate or the low energy cutoff. 

In some unique cases, RHESSI HXR observations show contributions from super-hot ($\rm T_{e}>30~MK$) coronal plasma, which are spatially and spectrally distinct from the hot plasma associated with chromospheric evaporation \citep{Caspi2010SuperhotPlasmaRHESSI}. Emission from these plasma may contribute to the thermal component of flare HXR spectra. However, the flares presented in this study have GOES temperatures of $\rm T_{GOES}\lesssim15~MK$ and RHESSI temperatures of $\rm T_{RHESSI}\lesssim20~MK$, therefore lacking a clear super-hot component. The temporal relationship between the \lya\ and HXR flux in Figure~\ref{fig:raw_lightcurves} further supports the assumption that the emission for these flares is primarily chromospheric in origin.  

Finally, the energy content of the nonthermal electrons in each flare was found to be approximately on the order of $10^{30}$~erg. Comparing this to the radiated energy in \lya, we were able to account for 2.0--7.9\% of the chromospheric energy budget in this line alone. These energies are found to be consistent with studies by \cite{Milligan2012EUVFlare, Milligan2014EUVEnergy} and \cite{deCosta2009NTLyaCorr}, the former of which found the \lya\ energy radiated during solar flares to be comparable to the total energy radiated across the entire EVE spectral range (50--1050\AA), further demonstrating the energetic significance of \lya\ as a single emission line.

For \ion{He}{2}, the peak enhancements were found to be larger than in \lya, ranging from 7.3--10.8\%. However, \ion{He}{2} did not demonstrate the same behavior as \lya\ in relation to the spectral index. The total energy radiated in \ion{He}{2} during the impulsive phase was found to equate to 0.6--1.4\% of the nonthermal electron energy, which was notably less than \lya\ for all three of the flares examined here. 

It has been shown that \lya\ emission can show dependence on flare magnitude (\citealt{Milligan2021B&CClassFlares}), however this study has demonstrated that the variability in \lya\ emission in solar flares may also depend on a more fundamental driver, such as the varying energetic properties of the nonthermal electrons. Recent studies have examined the effects of flare-induced radiation on the ionosphere across differing GOES classes (\citealt{Hayes2021Ionosphere, Barta2022IonosphericResponseFlares}). \cite{Milligan2020MXFlares} found that increased conductivity in the ionosphere may be attributed to increased \lya\ absorption in the E-region. Thus, irregularity in \lya\ emission between equivalent magnitude flares may impact our ability to predict ionospheric responses to solar flares based on their class, as well as our ability to extrapolate our understanding of the terrestrial impact of solar flares to our planetary neighbors (see \citealt{Yan2022EarthVenusMarsSolarFlares}).

On the relationship between nonthermal electrons and \lya\ emission, it is understood that the nonthermal electron parameters dictate the transport of energy into the chromosphere, and therefore the levels of chromospheric emission \citep{Kennedy2015RHDModellingFlare, Prochazka2018WLFElectronBeams}. Our findings imply that the level of flare-related \lya\ emission may be sensitive to the energetic properties of the nonthermal electrons incident on the chromosphere during solar flares. Care must be taken when drawing conclusions regarding the impact of individual parameters on the observed flare emission, as ultimately the energy distribution of the nonthermal electrons is dependent on the combination of the electron rate, spectral index, and low energy cutoff. Therefore, it may be useful to test a combination of parameters to discern the magnitude of their impact on chromospheric emission using radiative hydrodynamic simulation codes such as RADYN and RH (\citealt{Uitenbroek2001RH}). Synthetic emission profiles for both \lya\ and \ion{He}{2} may then be examined in an effort to validate the observational findings presented in this research.

For this study, calculations of the percentage contribution of \lya\ and \ion{He}{2} to the radiation of energy deposited in the chromosphere were founded on the assumption that their origin is entirely chromospheric. In the first instance, this would place the energy calculated in \lya\ at a lower limit, due to the significant optical-thickness of \lya\ in the chromosphere and the potential for self-absorption. With no spatially resolved observations of \lya\ for these flares, one must take great care when drawing conclusions regarding the energy attributed to chromospheric emission alone. We also note that our observations of \lya\ emission are taken across a relatively large spectral range given the broadband nature of the GOES/EUVS-E instrument. This means we cannot determine with absolute certainty that all of the emission measured for each flare is entirely attributable to the \lya\ line alone, as the E-channel spectral resolution also encapsulates the \ion{Si}{3} line at 1206\AA. We assume here that the irradiance data is dominated by the \lya\ emission, but it is impossible to confirm this exclusively from the observations presented in this study. For future study of \lya\ emission in solar flares, it would be of great value to have spatially and spectrally resolved measurements.

FISM is the most widely used model for providing estimates of solar emissions for periods of activity lacking observation. The first version of FISM saw application in numerous studies into the effect of solar radiation in both the terrestrial and martian atmospheres (\citealt{Qian2010FISMApplication, Qian2011FISMApplication, Qian2012FISMApplication, Lollo2012FISMApplicationMars}), as well as the development of the FISM-M model for the Extreme Ultraviolet Monitor (EUVM; \citealt{Eparvier2015MavenEUVM, Thiemann2017MavenEUVMModels}) onboard Mars Atmosphere and Volatile Experiment (MAVEN; \citealt{Jakosky2015MAVEN}). FISM2 offers significant improvements on the initial version of FISM, implementing more accurate data-sets, a greater number of flares, and longer time series. As of the release of FISM2, plans for improvements were already identified \citep{Chamberlin2020FISM}. Further upgrades to address the deficiencies of FISM2 model have been proposed for FISM3, including incorporating measurements and proxies from GOES/EUVS in order to improve the impulsive phase estimations, especially for optically thick emissions such as \lya\ and \ion{He}{2}. By including observations from GOES/EUVS-E and UARS/SOLSTICE, it may become possible to compute a CLV correction in the FUV wavelengths for the impulsive phase. Proposed improvements also plan to incorporate new measurements from \textit{Miniature X-ray Solar Spectrometer} and \textit{Dual-zone Aperture X-ray Solar Spectrometer} (MinXSS; DAXSS \citealt{Mason2016MinXSS, Woods2017MinXSS, Moore2018, Mason2019MinXSS, Schwab2020DAXSS}) to improve the soft X-ray emissions from 0.1-2.5 nm.

Finally, it is apparent that interest in studying flare-related \lya\ emission is growing, with newly launched and proposed missions featuring dedicated apparatus for observing at this wavelength. The new generation of GOES-R satellites feature a suite of co-observing SXR and \lya\ instruments as part of the \textit{EUV and X-ray Irradiance Sensors} (EXIS) which will provide pseudo \lya\ line profiles as part of its continuous observations for the next two decades (\citealt{Eparvier2009EUVS, Chamberlin2009XRS}). The recently launched Solar Orbiter mission includes a dedicated \lya\ channel as part of the \textit{High Resolution Imager} (HRI$\rm _{Ly\alpha}$) onboard the \textit{Extreme Ultraviolet Imager} (EUI), which is able to image the Sun in \lya\ at 1\arcsec\ resolution and 1s cadence (\citealt{Rochus2020EUI, Muller2020SolarOrbiter}). Future \lya\ missions also include \textit{Solar Spectral Irradiance Monitor} (SoSpIM) as part of the \textit{EUV High-Throughput Spectroscopic Telescope} (EUVST; \citealt{Shimizu2019EUVST}) onboard the Solar--C mission (\citealt{Watanabe2014SolarC}) and the \textit{\lya\ Solar Telescope} (LST) as part of the \textit{Advanced Space-based Solar Observatory} (ASO--S; \citealt{Li2019LST, Gan2023ASOS}). The findings presented here suggest that the spectral index, derived from HXR observations, may be an important tool for future development of models of \lya\ emission in solar flares. Instruments such as the \textit{Spectrometer Telescope for Imaging X-rays} (STIX; \citealt{Krucker2020STIX}) onboard Solar Orbiter and the \textit{Hard X-Ray Imager} (HXI; \citealt{Zhang2019HXI}) onboard ASO-S will provide further HXR observations of flares within Solar Cycle 25. The results presented in this study may guide interpretation of the observations taken by this new generation of instruments as part of future studies.

\begin{acknowledgments}

H.J.G would like to thank the UK’s Science \& Technology Facilities Council (ST/W507751/1) for supporting this research and for the award of Long Term Attachment funding to facilitate overseas fieldwork as part of this research. R.O.M would like to thank the UK’s Science \& Technologies Facilities Council for the award of an Ernest Rutherford Fellowship (ST/N004981/2). P.C.C would like to acknowledge support for this work by the SDO/EVE project at CU/LASP (PI Dr. Thomas Woods). H.J.G and R.O.M. would both like to thank Prof. Mihalis Mathioudakis for several stimulating discussions and initial feedback on this work. H.J.G would like to further extend his gratitude to Dr. Brian Dennis for his invaluable advice and guidance surrounding the analysis techniques presented in this paper. The authors would also like to thank the anonymous reviewer for their useful comments that contributed to improving this work.

\end{acknowledgments}

\section{Data Access Statement} \label{sec:access}

No new data was generated as part of this research. The pre-existing observational data used in this study are openly accessible through SolarSoftware (SSWIDL; \href{https://www.lmsal.com/solarsoft/}{https://www.lmsal.com/solarsoft/}). The FISM2 data are publicly available on the LISIRD data site on the FISM page, or directly here: \href{http://lasp.colorado.edu/lisird/}{http://lasp.colorado.edu/lisird/}.

\bibliography{references}{}

\begin{thebibliography}{}
\expandafter\ifx\csname natexlab\endcsname\relax\def\natexlab#1{#1}\fi
\providecommand{\url}[1]{\href{#1}{#1}}
\providecommand{\dodoi}[1]{doi:~\href{http://doi.org/#1}{\nolinkurl{#1}}}
\providecommand{\doeprint}[1]{\href{http://ascl.net/#1}{\nolinkurl{http://ascl.net/#1}}}
\providecommand{\doarXiv}[1]{\href{https://arxiv.org/abs/#1}{\nolinkurl{https://arxiv.org/abs/#1}}}

\bibitem[{{Allred} {et~al.}(2005){Allred}, {Hawley}, {Abbett}, \&
  {Carlsson}}]{Allred2005RADYN}
{Allred}, J.~C., {Hawley}, S.~L., {Abbett}, W.~P., \& {Carlsson}, M. 2005,
  \apj, 630, 573, \dodoi{10.1086/431751}

\bibitem[{Allred {et~al.}(2015)Allred, Kowalski, \& Carlsson}]{Allred_2015}
Allred, J.~C., Kowalski, A.~F., \& Carlsson, M. 2015, \apj, 809, 104,
  \dodoi{10.1088/0004-637x/809/1/104}

\bibitem[{{Auch{\`e}re}(2005)}]{Auchere2005LyaHeII}
{Auch{\`e}re}, F. 2005, \apj, 622, 737, \dodoi{10.1086/427903}

\bibitem[{{Baliukin} {et~al.}(2019){Baliukin}, {Bertaux}, {Qu{\'e}merais},
  {Izmodenov}, \& {Schmidt}}]{Baluikin2019Geocorona}
{Baliukin}, I.~I., {Bertaux}, J.~L., {Qu{\'e}merais}, E., {Izmodenov}, V.~V.,
  \& {Schmidt}, W. 2019, Journal of Geophysical Research (Space Physics), 124,
  861, \dodoi{10.1029/2018JA026136}

\bibitem[{Barta {et~al.}(2022)Barta, Natras, Srećković, Koronczay, Schmidt,
  \& Šulic}]{Barta2022IonosphericResponseFlares}
Barta, V., Natras, R., Srećković, V., {et~al.} 2022, Frontiers in
  Environmental Science, 10, \dodoi{10.3389/fenvs.2022.904335}

\bibitem[{Boerner {et~al.}(2012)Boerner, Edwards, Lemen, Rausch, Schrijver,
  Shine, Shing, Stern, Tarbell, Title, Wolfson, Soufli, Spiller, Gullikson,
  McKenzie, Windt, Golub, Podgorski, Testa, \& Weber}]{Boerner2012AIA}
Boerner, P., Edwards, C., Lemen, J., {et~al.} 2012, Solar Physics, 275, 41

\bibitem[{{Brekke} {et~al.}(1996){Brekke}, {Rottman}, {Fontenla}, \&
  {Judge}}]{Breke1996X3Flare}
{Brekke}, P., {Rottman}, G.~J., {Fontenla}, J., \& {Judge}, P.~G. 1996, \apj,
  468, 418, \dodoi{10.1086/177701}

\bibitem[{{Brown}(1971)}]{Brown1971CTTM}
{Brown}, J.~C. 1971, Solar Physics, 18, 489, \dodoi{10.1007/BF00149070}

\bibitem[{{Brown}(1972)}]{Brown1972CTTM}
---. 1972, Solar Physics, 26, 441, \dodoi{10.1007/BF00165286}

\bibitem[{{Brown}(1973)}]{Brown1973CTTM}
---. 1973, Solar Physics, 28, 151, \dodoi{10.1007/BF00152919}

\bibitem[{{Brown} {et~al.}(2018){Brown}, {Fletcher}, {Kerr}, {Labrosse},
  {Kowalski}, \& {De La Cruz Rodr{\'\i}guez}}]{Brown2018LymanModelling}
{Brown}, S.~A., {Fletcher}, L., {Kerr}, G.~S., {et~al.} 2018, \apj, 862, 59,
  \dodoi{10.3847/1538-4357/aacc29}

\bibitem[{Brown {et~al.}(2018)Brown, Fletcher, Kerr, Labrosse, Kowalski, \&
  Rodríguez}]{Brown2018RADYN_Lyman}
Brown, S.~A., Fletcher, L., Kerr, G.~S., {et~al.} 2018, The Astrophysical
  Journal, 862, 59, \dodoi{10.3847/1538-4357/aacc29}

\bibitem[{{Carmichael}(1964)}]{Carmichael1964CSHKP}
{Carmichael}, H. 1964, in NASA Special Publication, Vol.~50, 451

\bibitem[{Caspi \& Lin(2010)}]{Caspi2010SuperhotPlasmaRHESSI}
Caspi, A., \& Lin, R.~P. 2010, The Astrophysical Journal Letters, 725, L161,
  \dodoi{10.1088/2041-8205/725/2/L161}

\bibitem[{Chamberlin {et~al.}(2007)Chamberlin, Woods, \&
  Eparvier}]{Chamberlin2007FISM}
Chamberlin, P.~C., Woods, T.~N., \& Eparvier, F.~G. 2007, Space Weather, 5,
  \dodoi{https://doi.org/10.1029/2007SW000316}

\bibitem[{{Chamberlin} {et~al.}(2008){Chamberlin}, {Woods}, \&
  {Eparvier}}]{Chamberlin2008FISMoriginal}
{Chamberlin}, P.~C., {Woods}, T.~N., \& {Eparvier}, F.~G. 2008, Space Weather,
  6, S05001, \dodoi{10.1029/2007SW000372}

\bibitem[{{Chamberlin} {et~al.}(2009){Chamberlin}, {Woods}, {Eparvier}, \&
  {Jones}}]{Chamberlin2009XRS}
{Chamberlin}, P.~C., {Woods}, T.~N., {Eparvier}, F.~G., \& {Jones}, A.~R. 2009,
  in Society of Photo-Optical Instrumentation Engineers (SPIE) Conference
  Series, Vol. 7438, Solar Physics and Space Weather Instrumentation III, ed.
  S.~{Fineschi} \& J.~A. {Fennelly}, 743802, \dodoi{10.1117/12.826807}

\bibitem[{Chamberlin {et~al.}(2018)Chamberlin, Woods, Didkovsky, Eparvier,
  Jones, Machol, Mason, Snow, Thiemann, Viereck, \&
  Woodraska}]{Chamberlin2018SolarStorm}
Chamberlin, P.~C., Woods, T.~N., Didkovsky, L., {et~al.} 2018, Space Weather,
  16, 1470, \dodoi{https://doi.org/10.1029/2018SW001866}

\bibitem[{Chamberlin {et~al.}(2020)Chamberlin, Eparvier, Knoer, Leise,
  Pankratz, Snow, Templeman, Thiemann, Woodraska, \&
  Woods}]{Chamberlin2020FISM}
Chamberlin, P.~C., Eparvier, F.~G., Knoer, V., {et~al.} 2020, Space Weather,
  18, e2020SW002588, \dodoi{https://doi.org/10.1029/2020SW002588}

\bibitem[{Chubb {et~al.}(1957)Chubb, Friedman, Kreplin, \&
  Kupperian~Jr.}]{Chubb1957SoundingRockets}
Chubb, T.~A., Friedman, H., Kreplin, R.~W., \& Kupperian~Jr., J.~E. 1957,
  Journal of Geophysical Research (1896-1977), 62, 389,
  \dodoi{https://doi.org/10.1029/JZ062i003p00389}

\bibitem[{{Curdt} {et~al.}(2001){Curdt}, {Brekke}, {Feldman}, {Wilhelm},
  {Dwivedi}, {Sch{\"u}hle}, \& {Lemaire}}]{Curdt2001LyaBrightest}
{Curdt}, W., {Brekke}, P., {Feldman}, U., {et~al.} 2001, \aap, 375, 591,
  \dodoi{10.1051/0004-6361:20010364}

\bibitem[{{da Costa} {et~al.}(2009){da Costa}, {Fletcher}, {Labrosse}, \&
  {Zuccarello}}]{deCosta2009NTLyaCorr}
{da Costa}, F.~R., {Fletcher}, L., {Labrosse}, N., \& {Zuccarello}, F. 2009,
  A\&A, 507, 1005, \dodoi{10.1051/0004-6361/200912651}

\bibitem[{{Delaboudini{\`e}re} {et~al.}(1995){Delaboudini{\`e}re}, {Artzner},
  {Brunaud}, {Gabriel}, {Hochedez}, {Millier}, {Song}, {Au}, {Dere}, {Howard},
  {Kreplin}, {Michels}, {Moses}, {Defise}, {Jamar}, {Rochus}, {Chauvineau},
  {Marioge}, {Catura}, {Lemen}, {Shing}, {Stern}, {Gurman}, {Neupert},
  {Maucherat}, {Clette}, {Cugnon}, \& {Van Dessel}}]{Delaboudini1995SOHOEIT}
{Delaboudini{\`e}re}, J.~P., {Artzner}, G.~E., {Brunaud}, J., {et~al.} 1995,
  \solphys, 162, 291, \dodoi{10.1007/BF00733432}

\bibitem[{{Dennis} \& {Schwartz}(1989)}]{Dennis&Schwartz1982FlareBible}
{Dennis}, B.~R., \& {Schwartz}, R.~A. 1989, \solphys, 121, 75,
  \dodoi{10.1007/BF00161688}

\bibitem[{{Domingo} {et~al.}(1995){Domingo}, {Fleck}, \&
  {Poland}}]{Domingo1995SOHO}
{Domingo}, V., {Fleck}, B., \& {Poland}, A.~I. 1995, \solphys, 162, 1,
  \dodoi{10.1007/BF00733425}

\bibitem[{{Dominique} {et~al.}(2013){Dominique}, {Hochedez}, {Schmutz},
  {Dammasch}, {Shapiro}, {Kretzschmar}, {Zhukov}, {Gillotay}, {Stockman}, \&
  {BenMoussa}}]{Dominique2012LYRA}
{Dominique}, M., {Hochedez}, J.~F., {Schmutz}, W., {et~al.} 2013, \solphys,
  286, 21, \dodoi{10.1007/s11207-013-0252-5}

\bibitem[{{Dominique} {et~al.}(2018){Dominique}, {Zhukov}, {Heinzel},
  {Dammasch}, {Wauters}, {Dolla}, {Shestov}, {Kretzschmar}, {Machol},
  {Lapenta}, \& {Schmutz}}]{Dominique2018NTLyaCorr}
{Dominique}, M., {Zhukov}, A.~N., {Heinzel}, P., {et~al.} 2018, \apjl, 867,
  L24, \dodoi{10.3847/2041-8213/aaeace}

\bibitem[{Eparvier {et~al.}(2015)Eparvier, Chamberlin, Woods, \&
  Thiemann}]{Eparvier2015MavenEUVM}
Eparvier, F., Chamberlin, P., Woods, T., \& Thiemann, E. 2015, Space Science
  Reviews, 195, \dodoi{10.1007/s11214-015-0195-2}

\bibitem[{{Eparvier} {et~al.}(2009){Eparvier}, {Crotser}, {Jones},
  {McClintock}, {Snow}, \& {Woods}}]{Eparvier2009EUVS}
{Eparvier}, F.~G., {Crotser}, D., {Jones}, A.~R., {et~al.} 2009, in Society of
  Photo-Optical Instrumentation Engineers (SPIE) Conference Series, Vol. 7438,
  Solar Physics and Space Weather Instrumentation III, ed. S.~{Fineschi} \&
  J.~A. {Fennelly}, 743804, \dodoi{10.1117/12.826445}

\bibitem[{{Evans} {et~al.}(2010){Evans}, {Strickland}, {Woo}, {McMullin},
  {Plunkett}, {Viereck}, {Hill}, {Woods}, \& {Eparvier}}]{Evans2010EUVS}
{Evans}, J.~S., {Strickland}, D.~J., {Woo}, W.~K., {et~al.} 2010, \solphys,
  262, 71, \dodoi{10.1007/s11207-009-9491-x}

\bibitem[{{Freeland} \& {Handy}(1998)}]{Freeland1998SSWIDL}
{Freeland}, S.~L., \& {Handy}, B.~N. 1998, \solphys, 182, 497,
  \dodoi{10.1023/A:1005038224881}

\bibitem[{{Gan} {et~al.}(2023){Gan}, {Zhu}, {Deng}, {Zhang}, {Chen}, {Huang},
  {Deng}, {Wu}, {Zhang}, {Li}, {Su}, {Su}, {Feng}, {Wu}, {Cui}, {Wang},
  {Chang}, {Yin}, {Xiong}, {Chen}, {Yang}, {Li}, {Lin}, {Hou}, {Bai}, {Chen},
  {Zhang}, {Hu}, {Liang}, {Wang}, {Song}, {Guo}, {He}, {Zhang}, {Wang}, {Bao},
  {Cao}, {Bai}, {Chen}, {He}, {Li}, {Zhang}, {Liao}, {Jiang}, {Li}, {Su},
  {Lei}, {Chen}, {Li}, {Zhao}, {Li}, {Ge}, {Zou}, {Hu}, {Su}, {Ji}, {Gu},
  {Zheng}, {Xu}, \& {Wang}}]{Gan2023ASOS}
{Gan}, W., {Zhu}, C., {Deng}, Y., {et~al.} 2023, \solphys, 298, 68,
  \dodoi{10.1007/s11207-023-02166-x}

\bibitem[{Garcia {et~al.}(2007)Garcia, Marsh, Kinnison, Boville, \&
  Sassi}]{Garcia2007WACCMX}
Garcia, R.~R., Marsh, D.~R., Kinnison, D.~E., Boville, B.~A., \& Sassi, F.
  2007, Journal of Geophysical Research: Atmospheres, 112,
  \dodoi{https://doi.org/10.1029/2006JD007485}

\bibitem[{{Gordino} {et~al.}(2022){Gordino}, {Auch{\`e}re}, {Vial},
  {Bocchialini}, {Hassler}, {Bando}, {Ishikawa}, {Kano}, {Kobayashi},
  {Narukage}, {Trujillo Bueno}, \& {Winebarger}}]{Gordino2022LyaHeII}
{Gordino}, M., {Auch{\`e}re}, F., {Vial}, J.~C., {et~al.} 2022, \aap, 657, A86,
  \dodoi{10.1051/0004-6361/202141960}

\bibitem[{{Grigis} \& {Benz}(2004)}]{GrigisBenze2004SHS}
{Grigis}, P.~C., \& {Benz}, A.~O. 2004, \aap, 426, 1093,
  \dodoi{10.1051/0004-6361:20041367}

\bibitem[{Hanser \& Sellers(1996)}]{Hanser1996GOES_XRS}
Hanser, F.~A., \& Sellers, F.~B. 1996, in GOES-8 and Beyond, ed. E.~R.
  Washwell, Vol. 2812, International Society for Optics and Photonics (SPIE),
  344 -- 352, \dodoi{10.1117/12.254082}

\bibitem[{Hayes {et~al.}(2021)Hayes, O’Hara, Murray, \&
  Gallagher}]{Hayes2021Ionosphere}
Hayes, L.~A., O’Hara, O. S.~D., Murray, S.~A., \& Gallagher, P.~T. 2021,
  Solar Physics, 296, \dodoi{10.1007/s11207-021-01898-y}

\bibitem[{{Hirayama}(1974)}]{Hiryama1974CSHKP}
{Hirayama}, T. 1974, \solphys, 34, 323, \dodoi{10.1007/BF00153671}

\bibitem[{{Holman} {et~al.}(2003){Holman}, {Sui}, {Schwartz}, \&
  {Emslie}}]{Holman2003ElectronBremmHXRSpectra}
{Holman}, G.~D., {Sui}, L., {Schwartz}, R.~A., \& {Emslie}, A.~G. 2003, \apjl,
  595, L97, \dodoi{10.1086/378488}

\bibitem[{{Hudson} {et~al.}(1992){Hudson}, {Acton}, {Hirayama}, \&
  {Uchida}}]{Hudson1992YokkohWLF}
{Hudson}, H.~S., {Acton}, L.~W., {Hirayama}, T., \& {Uchida}, Y. 1992, \pasj,
  44, L77

\bibitem[{{Hudson} {et~al.}(2006){Hudson}, {Wolfson}, \&
  {Metcalf}}]{Hudson2006WLFTRACE}
{Hudson}, H.~S., {Wolfson}, C.~J., \& {Metcalf}, T.~R. 2006, \solphys, 234, 79,
  \dodoi{10.1007/s11207-006-0056-y}

\bibitem[{{Ireland} {et~al.}(2013){Ireland}, {Tolbert}, {Schwartz}, {Holman},
  \& {Dennis}}]{Ireland2013ModellingHXRFlares}
{Ireland}, J., {Tolbert}, A.~K., {Schwartz}, R.~A., {Holman}, G.~D., \&
  {Dennis}, B.~R. 2013, \apj, 769, 89, \dodoi{10.1088/0004-637X/769/2/89}

\bibitem[{{Jakosky} {et~al.}(2015){Jakosky}, {Lin}, {Grebowsky}, {Luhmann},
  {Mitchell}, {Beutelschies}, {Priser}, {Acuna}, {Andersson}, {Baird}, {Baker},
  {Bartlett}, {Benna}, {Bougher}, {Brain}, {Carson}, {Cauffman}, {Chamberlin},
  {Chaufray}, {Cheatom}, {Clarke}, {Connerney}, {Cravens}, {Curtis}, {Delory},
  {Demcak}, {DeWolfe}, {Eparvier}, {Ergun}, {Eriksson}, {Espley}, {Fang},
  {Folta}, {Fox}, {Gomez-Rosa}, {Habenicht}, {Halekas}, {Holsclaw}, {Houghton},
  {Howard}, {Jarosz}, {Jedrich}, {Johnson}, {Kasprzak}, {Kelley}, {King},
  {Lankton}, {Larson}, {Leblanc}, {Lefevre}, {Lillis}, {Mahaffy}, {Mazelle},
  {McClintock}, {McFadden}, {Mitchell}, {Montmessin}, {Morrissey}, {Peterson},
  {Possel}, {Sauvaud}, {Schneider}, {Sidney}, {Sparacino}, {Stewart}, {Tolson},
  {Toublanc}, {Waters}, {Woods}, {Yelle}, \& {Zurek}}]{Jakosky2015MAVEN}
{Jakosky}, B.~M., {Lin}, R.~P., {Grebowsky}, J.~M., {et~al.} 2015, \ssr, 195,
  3, \dodoi{10.1007/s11214-015-0139-x}

\bibitem[{{Jing} {et~al.}(2020){Jing}, {Pan}, {Yang}, {Song}, {Tian}, {Li},
  {Cheng}, {Hong}, \& {Ding}}]{Jing2020CoolingLyaEmission}
{Jing}, Z., {Pan}, W., {Yang}, Y., {et~al.} 2020, \apj, 904, 41,
  \dodoi{10.3847/1538-4357/abbacc}

\bibitem[{{Kennedy} {et~al.}(2015){Kennedy}, {Milligan}, {Allred},
  {Mathioudakis}, \& {Keenan}}]{Kennedy2015RHDModellingFlare}
{Kennedy}, M.~B., {Milligan}, R.~O., {Allred}, J.~C., {Mathioudakis}, M., \&
  {Keenan}, F.~P. 2015, \aap, 578, A72, \dodoi{10.1051/0004-6361/201425144}

\bibitem[{Kleint {et~al.}(2016)Kleint, Heinzel, Judge, \&
  Krucker}]{Kleint2016ContinuumEnergy}
Kleint, L., Heinzel, P., Judge, P., \& Krucker, S. 2016, The Astrophysical
  Journal, 816, 88, \dodoi{10.3847/0004-637x/816/2/88}

\bibitem[{{Kontar} {et~al.}(2006){Kontar}, {MacKinnon}, {Schwartz}, \&
  {Brown}}]{Kontar2006AlbedoBackScatter}
{Kontar}, E.~P., {MacKinnon}, A.~L., {Schwartz}, R.~A., \& {Brown}, J.~C. 2006,
  \aap, 446, 1157, \dodoi{10.1051/0004-6361:20053672}

\bibitem[{{Kopp} \& {Pneuman}(1976)}]{KoppPneuman1976CSHKP}
{Kopp}, R.~A., \& {Pneuman}, G.~W. 1976, \solphys, 50, 85,
  \dodoi{10.1007/BF00206193}

\bibitem[{{Kretzschmar}(2011)}]{Kretzschmar2011TSI}
{Kretzschmar}, M. 2011, \aap, 530, A84, \dodoi{10.1051/0004-6361/201015930}

\bibitem[{{Kretzschmar} {et~al.}(2010){Kretzschmar}, {de Wit}, {Schmutz},
  {Mekaoui}, {Hochedez}, \& {Dewitte}}]{Kretzschmar2010TSI}
{Kretzschmar}, M., {de Wit}, T.~D., {Schmutz}, W., {et~al.} 2010, Nature
  Physics, 6, 690, \dodoi{10.1038/nphys1741}

\bibitem[{Kretzschmar {et~al.}(2012)Kretzschmar, Dominique, \&
  Dammasch}]{Kretzschmar2012LyaProbaLYRA}
Kretzschmar, M., Dominique, M., \& Dammasch, I.~E. 2012, Solar Physics, 286,
  221, \dodoi{10.1007/s11207-012-0175-6}

\bibitem[{{Krucker} {et~al.}(2020){Krucker}, {Hurford, G. J.}, {Grimm, O.},
  {K\"ogl, S.}, {Gr\"obelbauer, H.-P}, {Etesi, L.}, {Casadei, D.}, {Csillaghy,
  A.}, {Benz, A. O.}, {Arnold, N. G.}, {Molendini, F.}, {Orleanski, P.},
  {Schori, D.}, {Xiao, H.}, {Kuhar, M.}, {Hochmuth, N.}, {Felix, S.},
  {Schramka, F.}, {Marcin, S.}, {Kobler, S.}, {Iseli, L.}, {Dreier, M.},
  {Wiehl, H. J.}, {Kleint, L.}, {Battaglia, M.}, {Lastufka, E.}, {Sathiapal,
  H.}, {Lapadula, K.}, {Bednarzik, M.}, {Birrer, G.}, {Stutz, St.}, {Wild,
  Ch.}, {Marone, F.}, {Skup, K. R.}, {Cichocki, A.}, {Ber, K.}, {Rutkowski,
  K.}, {Bujwan, W.}, {Juchnikowski, G.}, {Winkler, M.}, {Darmetko, M.},
  {Michalska, M.}, {Seweryn, K.}, {Bialek, A.}, {Osica, P.}, {Sylwester, J.},
  {Kowalinski, M.}, {\'{}Scislowski, D.}, {Siarkowski, M.}, {Ste\'{}slicki,
  M.}, {Mrozek, T.}, {Podg\'orski, P.}, {Meuris, A.}, {Limousin, O.}, {Gevin,
  O.}, {Le Mer, I.}, {Brun, S.}, {Strugarek, A.}, {Vilmer, N.}, {Musset, S.},
  {Maksimovi\'{}c, M.}, {F\'arn\'{\i}k, F.}, {Koz\'acek, Z.}, {Kasparov\'a,
  J.}, {Mann, G.}, {\"Onel, H.}, {Warmuth, A.}, {Rendtel, J.}, {Anderson, J.},
  {Bauer, S.}, {Dionies, F.}, {Paschke, J.}, {Pl\"uschke, D.}, {Woche, M.},
  {Schuller, F.}, {Veronig, A. M.}, {Dickson, E. C. M.}, {Gallagher, P. T.},
  {Maloney, S. A.}, {Bloomfield, D. S.}, {Piana, M.}, {Massone, A. M.},
  {Benvenuto, F.}, {Massa, P.}, {Schwartz, R. A.}, {Dennis, B. R.}, {van Beek,
  H. F.}, {Rodr\'{\i}guez-Pacheco, J.}, \& {Lin, R. P.}}]{Krucker2020STIX}
{Krucker}, S., {Hurford, G. J.}, {Grimm, O.}, {et~al.} 2020, A\&A, 642, A15,
  \dodoi{10.1051/0004-6361/201937362}

\bibitem[{{Kumar} \& {Kumar}(2018)}]{Kumar2018D-RegionSXR}
{Kumar}, A., \& {Kumar}, S. 2018, Earth, Planets and Space, 70, 29,
  \dodoi{10.1186/s40623-018-0794-8}

\bibitem[{{Li} {et~al.}(2019){Li}, {Chen}, {Feng}, {Li}, {Huang}, {Li}, {Lu},
  {Xue}, {Ying}, {Zhao}, {Yang}, {Gan}, {Fang}, {Song}, {Wang}, {Guo}, {He},
  {Zhu}, {Zhu}, {Deng}, {Bao}, {Cao}, \& {Yang}}]{Li2019LST}
{Li}, H., {Chen}, B., {Feng}, L., {et~al.} 2019, Research in Astronomy and
  Astrophysics, 19, 158, \dodoi{10.1088/1674-4527/19/11/158}

\bibitem[{Li {et~al.}(2022)Li, Li, Song, Battaglia, Xiao, Krucker, Schühle,
  Li, Gan, \& Ding}]{Li2022LyaNTCorr}
Li, Y., Li, Q., Song, D.-C., {et~al.} 2022, The Astrophysical Journal, 936,
  142, \dodoi{10.3847/1538-4357/ac897c}

\bibitem[{{Lin} {et~al.}(2002){Lin}, {Dennis}, {Hurford}, {Smith}, {Zehnder},
  {Harvey}, {Curtis}, {Pankow}, {Turin}, {Bester}, {Csillaghy}, {Lewis},
  {Madden}, {van Beek}, {Appleby}, {Raudorf}, {McTiernan}, {Ramaty}, {Schmahl},
  {Schwartz}, {Krucker}, {Abiad}, {Quinn}, {Berg}, {Hashii}, {Sterling},
  {Jackson}, {Pratt}, {Campbell}, {Malone}, {Landis}, {Barrington-Leigh},
  {Slassi-Sennou}, {Cork}, {Clark}, {Amato}, {Orwig}, {Boyle}, {Banks},
  {Shirey}, {Tolbert}, {Zarro}, {Snow}, {Thomsen}, {Henneck}, {McHedlishvili},
  {Ming}, {Fivian}, {Jordan}, {Wanner}, {Crubb}, {Preble}, {Matranga}, {Benz},
  {Hudson}, {Canfield}, {Holman}, {Crannell}, {Kosugi}, {Emslie}, {Vilmer},
  {Brown}, {Johns-Krull}, {Aschwanden}, {Metcalf}, \& {Conway}}]{Lin2002RHESSI}
{Lin}, R.~P., {Dennis}, B.~R., {Hurford}, G.~J., {et~al.} 2002, \solphys, 210,
  3, \dodoi{10.1023/A:1022428818870}

\bibitem[{Lollo {et~al.}(2012)Lollo, Withers, Fallows, Girazian, Matta, \&
  Chamberlin}]{Lollo2012FISMApplicationMars}
Lollo, A., Withers, P., Fallows, K., {et~al.} 2012, Journal of Geophysical
  Research: Space Physics, 117, \dodoi{https://doi.org/10.1029/2011JA017399}

\bibitem[{{Lu} {et~al.}(2021){Lu}, {Feng}, {Li}, {Ying}, {Li}, {Gan}, {Li}, \&
  {Zhou}}]{Lu2021LymanNTCorr}
{Lu}, L., {Feng}, L., {Li}, D., {et~al.} 2021, \apjs, 253, 29,
  \dodoi{10.3847/1538-4365/abd79b}

\bibitem[{Marsh {et~al.}(2013)Marsh, Mills, Kinnison, Lamarque, Calvo, \&
  Polvani}]{Marsh2013WACCM}
Marsh, D., Mills, M., Kinnison, D., {et~al.} 2013, Journal Of Climate, 26,
  7372, \dodoi{10.1175/JCLI-D-12-00558.1}

\bibitem[{{Mason} {et~al.}(2020){Mason}, {Woods}, {Chamberlin}, {Jones},
  {Kohnert}, {Schwab}, {Sewell}, {Caspi}, {Moore}, {Palo}, {Solomon}, \&
  {Warren}}]{Mason2019MinXSS}
{Mason}, J., {Woods}, T., {Chamberlin}, P., {et~al.} 2020, Advances in Space
  Research, 66, 3, \dodoi{10.1016/j.asr.2019.02.011}

\bibitem[{{Mason} {et~al.}(2016){Mason}, {Woods}, {Caspi}, {Chamberlin},
  {Moore}, {Jones}, {Kohnert}, {Li}, {Palo}, \& {Solomon}}]{Mason2016MinXSS}
{Mason}, J.~P., {Woods}, T.~N., {Caspi}, A., {et~al.} 2016, Journal of
  Spacecraft and Rockets, 53, 328, \dodoi{10.2514/1.A33351}

\bibitem[{{McClintock} {et~al.}(2005){McClintock}, {Rottman}, \&
  {Woods}}]{McClintock2005SOLSTICE}
{McClintock}, W.~E., {Rottman}, G.~J., \& {Woods}, T.~N. 2005, \solphys, 230,
  225, \dodoi{10.1007/s11207-005-7432-x}

\bibitem[{{McRae} \& {Thomson}(2004)}]{McRae2004D-RegionSXR}
{McRae}, W.~M., \& {Thomson}, N.~R. 2004, Journal of Atmospheric and
  Solar-Terrestrial Physics, 66, 77, \dodoi{10.1016/j.jastp.2003.09.009}

\bibitem[{Meier \& Prinz(1970)}]{Meier1976Geocorona}
Meier, R.~R., \& Prinz, D.~K. 1970, Journal of Geophysical Research
  (1896-1977), 75, 6969, \dodoi{https://doi.org/10.1029/JA075i034p06969}

\bibitem[{{Milligan}(2015)}]{Milligan2015EUVInvitedReview}
{Milligan}, R.~O. 2015, Solar Physics, 290, 3399,
  \dodoi{10.1007/s11207-015-0748-2}

\bibitem[{Milligan(2021)}]{Milligan2021B&CClassFlares}
Milligan, R.~O. 2021, Solar Physics, 296, \dodoi{10.1007/s11207-021-01796-3}

\bibitem[{{Milligan} \& {Chamberlin}(2016)}]{Milligan2016SDO_EVE_MEGSP}
{Milligan}, R.~O., \& {Chamberlin}, P.~C. 2016, \aap, 587, A123,
  \dodoi{10.1051/0004-6361/201526682}

\bibitem[{Milligan {et~al.}(2012)Milligan, Chamberlin, Hudson, Woods,
  Mathioudakis, Fletcher, Kowalski, \& Keenan}]{Milligan2012EUVFlare}
Milligan, R.~O., Chamberlin, P.~C., Hudson, H.~S., {et~al.} 2012, \apj, 748,
  L14, \dodoi{10.1088/2041-8205/748/1/l14}

\bibitem[{Milligan {et~al.}(2017)Milligan, Fleck, Ireland, Fletcher, \&
  Dennis}]{Milligan2017NTLyaCorr}
Milligan, R.~O., Fleck, B., Ireland, J., Fletcher, L., \& Dennis, B.~R. 2017,
  The Astrophysical Journal, 848, L8, \dodoi{10.3847/2041-8213/aa8f3a}

\bibitem[{Milligan {et~al.}(2020)Milligan, Hudson, Chamberlin, Hannah, \&
  Hayes}]{Milligan2020MXFlares}
Milligan, R.~O., Hudson, H.~S., Chamberlin, P.~C., Hannah, I.~G., \& Hayes,
  L.~A. 2020, Space Weather, 18, \dodoi{10.1029/2019sw002331}

\bibitem[{Milligan {et~al.}(2014)Milligan, Kerr, Dennis, Hudson, Fletcher,
  Allred, Chamberlin, Ireland, Mathioudakis, \& Keenan}]{Milligan2014EUVEnergy}
Milligan, R.~O., Kerr, G.~S., Dennis, B.~R., {et~al.} 2014, \apj, 793, 70,
  \dodoi{10.1088/0004-637x/793/2/70}

\bibitem[{{Moore} {et~al.}(2018){Moore}, {Caspi}, {Woods}, {Chamberlin},
  {Dennis}, {Jones}, {Mason}, {Schwartz}, \& {Tolbert}}]{Moore2018}
{Moore}, C.~S., {Caspi}, A., {Woods}, T.~N., {et~al.} 2018, Solar Physics, 293,
  21, \dodoi{10.1007/s11207-018-1243-3}

\bibitem[{{M\"uller} {et~al.}(2020){M\"uller}, {St. Cyr, O. C.}, {Zouganelis,
  I.}, {Gilbert, H. R.}, {Marsden, R.}, {Nieves-Chinchilla, T.}, {Antonucci,
  E.}, {Auch\`ere, F.}, {Berghmans, D.}, {Horbury, T. S.}, {Howard, R. A.},
  {Krucker, S.}, {Maksimovic, M.}, {Owen, C. J.}, {Rochus, P.},
  {Rodriguez-Pacheco, J.}, {Romoli, M.}, {Solanki, S. K.}, {Bruno, R.},
  {Carlsson, M.}, {Fludra, A.}, {Harra, L.}, {Hassler, D. M.}, {Livi, S.},
  {Louarn, P.}, {Peter, H.}, {Sch\"uhle, U.}, {Teriaca, L.}, {del Toro Iniesta,
  J. C.}, {Wimmer-Schweingruber, R. F.}, {Marsch, E.}, {Velli, M.}, {De Groof,
  A.}, {Walsh, A.}, \& {Williams, D.}}]{Muller2020SolarOrbiter}
{M\"uller}, D., {St. Cyr, O. C.}, {Zouganelis, I.}, {et~al.} 2020, A\&A, 642,
  A1, \dodoi{10.1051/0004-6361/202038467}

\bibitem[{Neale {et~al.}(2013)Neale, Richter, Park, Lauritzen, Vavrus, Rasch,
  \& Zhang}]{Neale2013WACCMX}
Neale, R.~B., Richter, J., Park, S., {et~al.} 2013, Journal of Climate, 26,
  5150 , \dodoi{10.1175/JCLI-D-12-00236.1}

\bibitem[{{Neidig}(1989)}]{Neidig1989WLF}
{Neidig}, D.~F. 1989, \solphys, 121, 261, \dodoi{10.1007/BF00161699}

\bibitem[{{Nina} {et~al.}(2018){Nina}, {{\v{C}}ade{\v{z}}},
  {Baj{\v{c}}eti{\'c}}, {Mitrovi{\'c}}, \& {Popovi{\'c}}}]{Nina2018D-RegionSXR}
{Nina}, A., {{\v{C}}ade{\v{z}}}, V.~M., {Baj{\v{c}}eti{\'c}}, J.,
  {Mitrovi{\'c}}, S.~T., \& {Popovi{\'c}}, L.~{\v{C}}. 2018, \solphys, 293, 64,
  \dodoi{10.1007/s11207-018-1279-4}

\bibitem[{{Nusinov} {et~al.}(2006){Nusinov}, {Kazachevskaya}, {Kuznetsov},
  {Myagkova}, \& {Yushkov}}]{Nusinov2006LyaNTCorr}
{Nusinov}, A.~A., {Kazachevskaya}, T.~V., {Kuznetsov}, S.~N., {Myagkova},
  I.~N., \& {Yushkov}, B.~Y. 2006, Solar System Research, 40, 282,
  \dodoi{10.1134/S0038094606040034}

\bibitem[{{Pesnell} {et~al.}(2012){Pesnell}, {Thompson}, \&
  {Chamberlin}}]{Pesnell2012SDOMain}
{Pesnell}, W.~D., {Thompson}, B.~J., \& {Chamberlin}, P.~C. 2012, \solphys,
  275, 3, \dodoi{10.1007/s11207-011-9841-3}

\bibitem[{{Proch{\'a}zka} {et~al.}(2018){Proch{\'a}zka}, {Reid}, {Milligan},
  {Sim{\~o}es}, {Allred}, \& {Mathioudakis}}]{Prochazka2018WLFElectronBeams}
{Proch{\'a}zka}, O., {Reid}, A., {Milligan}, R.~O., {et~al.} 2018, \apj, 862,
  76, \dodoi{10.3847/1538-4357/aaca37}

\bibitem[{Qian {et~al.}(2010)Qian, Burns, Chamberlin, \&
  Solomon}]{Qian2010FISMApplication}
Qian, L., Burns, A.~G., Chamberlin, P.~C., \& Solomon, S.~C. 2010, Journal of
  Geophysical Research: Space Physics, 115,
  \dodoi{https://doi.org/10.1029/2009JA015225}

\bibitem[{Qian {et~al.}(2011)Qian, Burns, Chamberlin, \&
  Solomon}]{Qian2011FISMApplication}
---. 2011, Journal of Geophysical Research: Space Physics, 116,
  \dodoi{https://doi.org/10.1029/2011JA016777}

\bibitem[{Qian {et~al.}(2012)Qian, Burns, Solomon, \&
  Chamberlin}]{Qian2012FISMApplication}
Qian, L., Burns, A.~G., Solomon, S.~C., \& Chamberlin, P.~C. 2012, Geophysical
  Research Letters, 39, \dodoi{https://doi.org/10.1029/2012GL051102}

\bibitem[{Qian {et~al.}(2018)Qian, Burns, Solomon, Smith, McInerney, Hunt,
  Marsh, Liu, Mlynczak, \& Vitt}]{Qian2018WACCMX}
Qian, L., Burns, A.~G., Solomon, S.~S., {et~al.} 2018, Journal of Geophysical
  Research: Space Physics, 123, 1006,
  \dodoi{https://doi.org/10.1002/2017JA024998}

\bibitem[{Raulin {et~al.}(2013)Raulin, Trottet, Kretzschmar, Macotela, Pacini,
  Bertoni, \& Dammasch}]{Raulin2013LyaIonosphere}
Raulin, J.-P., Trottet, G., Kretzschmar, M., {et~al.} 2013, Journal of
  Geophysical Research: Space Physics, 118, 570,
  \dodoi{https://doi.org/10.1029/2012JA017916}

\bibitem[{{Rochus} {et~al.}(2020){Rochus}, {Auch\`ere, F.}, {Berghmans, D.},
  {Harra, L.}, {Schmutz, W.}, {Sch\"uhle, U.}, {Addison, P.}, {Appourchaux,
  T.}, {Aznar Cuadrado, R.}, {Baker, D.}, {Barbay, J.}, {Bates, D.},
  {BenMoussa, A.}, {Bergmann, M.}, {Beurthe, C.}, {Borgo, B.}, {Bonte, K.},
  {Bouzit, M.}, {Bradley, L.}, {B\"uchel, V.}, {Buchlin, E.}, {B\"uchner, J.},
  {Cab\'e, F.}, {Cadiergues, L.}, {Chaigneau, M.}, {Chares, B.}, {Choque
  Cortez, C.}, {Coker, P.}, {Condamin, M.}, {Coumar, S.}, {Curdt, W.}, {Cutler,
  J.}, {Davies, D.}, {Davison, G.}, {Defise, J.-M.}, {Del Zanna, G.},
  {Delmotte, F.}, {Delouille, V.}, {Dolla, L.}, {Dumesnil, C.}, {D\"urig, F.},
  {Enge, R.}, {Fran\c{c}ois, S.}, {Fourmond, J.-J.}, {Gillis, J.-M.},
  {Giordanengo, B.}, {Gissot, S.}, {Green, L. M.}, {Guerreiro, N.}, {Guilbaud,
  A.}, {Gyo, M.}, {Haberreiter, M.}, {Hafiz, A.}, {Hailey, M.}, {Halain,
  J.-P.}, {Hansotte, J.}, {Hecquet, C.}, {Heerlein, K.}, {Hellin, M.-L.},
  {Hemsley, S.}, {Hermans, A.}, {Hervier, V.}, {Hochedez, J.-F.}, {Houbrechts,
  Y.}, {Ihsan, K.}, {Jacques, L.}, {J\'er\^ome, A.}, {Jones, J.}, {Kahle, M.},
  {Kennedy, T.}, {Klaproth, M.}, {Kolleck, M.}, {Koller, S.}, {Kotsialos, E.},
  {Kraaikamp, E.}, {Langer, P.}, {Lawrenson, A.}, {Le Clech\'{}, J.-C.},
  {Lenaerts, C.}, {Liebecq, S.}, {Linder, D.}, {Long, D. M.}, {Mampaey, B.},
  {Markiewicz-Innes, D.}, {Marquet, B.}, {Marsch, E.}, {Matthews, S.}, {Mazy,
  E.}, {Mazzoli, A.}, {Meining, S.}, {Meltchakov, E.}, {Mercier, R.}, {Meyer,
  S.}, {Monecke, M.}, {Monfort, F.}, {Morinaud, G.}, {Moron, F.}, {Mountney,
  L.}, {M\"uller, R.}, {Nicula, B.}, {Parenti, S.}, {Peter, H.}, {Pfiffner,
  D.}, {Philippon, A.}, {Phillips, I.}, {Plesseria, J.-Y.}, {Pylyser, E.},
  {Rabecki, F.}, {Ravet-Krill, M.-F.}, {Rebellato, J.}, {Renotte, E.},
  {Rodriguez, L.}, {Roose, S.}, {Rosin, J.}, {Rossi, L.}, {Roth, P.},
  {Rouesnel, F.}, {Roulliay, M.}, {Rousseau, A.}, {Ruane, K.}, {Scanlan, J.},
  {Schlatter, P.}, {Seaton, D. B.}, {Silliman, K.}, {Smit, S.}, {Smith, P. J.},
  {Solanki, S. K.}, {Spescha, M.}, {Spencer, A.}, {Stegen, K.}, {Stockman, Y.},
  {Szwec, N.}, {Tamiatto, C.}, {Tandy, J.}, {Teriaca, L.}, {Theobald, C.},
  {Tychon, I.}, {van Driel-Gesztelyi, L.}, {Verbeeck, C.}, {Vial, J.-C.},
  {Werner, S.}, {West, M. J.}, {Westwood, D.}, {Wiegelmann, T.}, {Willis, G.},
  {Winter, B.}, {Zerr, A.}, {Zhang, X.}, \& {Zhukov, A. N.}}]{Rochus2020EUI}
{Rochus}, P., {Auch\`ere, F.}, {Berghmans, D.}, {et~al.} 2020, A\&A, 642, A8,
  \dodoi{10.1051/0004-6361/201936663}

\bibitem[{{Rottman} {et~al.}(1993){Rottman}, {Woods}, \&
  {Sparn}}]{Rottman1993UARSPaper1}
{Rottman}, G.~J., {Woods}, T.~N., \& {Sparn}, T.~P. 1993, \jgr, 98, 10,667,
  \dodoi{10.1029/93JD00462}

\bibitem[{Saint-Hilaire \& Benz(2005)}]{stHilaire2005FlareEnergy}
Saint-Hilaire, P., \& Benz, A. 2005, A\&A, 435,
  \dodoi{10.1051/0004-6361:20041918}

\bibitem[{{Santandrea} {et~al.}(2013){Santandrea}, {Gantois}, {Strauch},
  {Teston}, {Tilmans}, {Baijot}, {Gerrits}, {De Groof}, {Schwehm}, \&
  {Zender}}]{Santandrea2013PROBA2}
{Santandrea}, S., {Gantois}, K., {Strauch}, K., {et~al.} 2013, \solphys, 286,
  5, \dodoi{10.1007/s11207-013-0289-5}

\bibitem[{{Schwab} {et~al.}(2020){Schwab}, {Sewell}, {Woods}, {Caspi}, {Mason},
  \& {Moore}}]{Schwab2020DAXSS}
{Schwab}, B.~D., {Sewell}, R. H.~A., {Woods}, T.~N., {et~al.} 2020, \apj, 904,
  20, \dodoi{10.3847/1538-4357/abba2a}

\bibitem[{{Shimizu} {et~al.}(2019){Shimizu}, {Imada}, {Kawate}, {Ichimoto},
  {Suematsu}, {Hara}, {Katsukawa}, {Kubo}, {Toriumi}, {Watanabe}, {Yokoyama},
  {Korendyke}, {Warren}, {Tarbell}, {De Pontieu}, {Teriaca}, {Sch{\"u}hle},
  {Solanki}, {Harra}, {Matthews}, {Fludra}, {Auch{\`e}re}, {Andretta},
  {Naletto}, \& {Zhukov}}]{Shimizu2019EUVST}
{Shimizu}, T., {Imada}, S., {Kawate}, T., {et~al.} 2019, in Society of
  Photo-Optical Instrumentation Engineers (SPIE) Conference Series, Vol. 11118,
  UV, X-Ray, and Gamma-Ray Space Instrumentation for Astronomy XXI, ed. O.~H.
  {Siegmund}, 1111807, \dodoi{10.1117/12.2528240}

\bibitem[{{Sim{\~o}es} {et~al.}(2016){Sim{\~o}es}, {Fletcher}, {Labrosse}, \&
  {Kerr}}]{Simoes2016HeII}
{Sim{\~o}es}, P.~J.~A., {Fletcher}, L., {Labrosse}, N., \& {Kerr}, G.~S. 2016,
  in Astronomical Society of the Pacific Conference Series, Vol. 504, Coimbra
  Solar Physics Meeting: Ground-based Solar Observations in the Space
  Instrumentation Era, ed. I.~{Dorotovic}, C.~E. {Fischer}, \& M.~{Temmer},
  197.
\newblock \doarXiv{1512.03477}

\bibitem[{Solomon {et~al.}(2018)Solomon, Liu, Marsh, McInerney, Qian, \&
  Vitt}]{Solomon2018WACCMX}
Solomon, S.~C., Liu, H.-L., Marsh, D.~R., {et~al.} 2018, Geophysical Research
  Letters, 45, 1567, \dodoi{https://doi.org/10.1002/2017GL076950}

\bibitem[{{Sturrock}(1966)}]{Sturrock1966CSHKP}
{Sturrock}, P.~A. 1966, \nat, 211, 695, \dodoi{10.1038/211695a0}

\bibitem[{Thiemann {et~al.}(2017)Thiemann, Chamberlin, Eparvier, Templeman,
  Woods, Bougher, \& Jakosky}]{Thiemann2017MavenEUVMModels}
Thiemann, E. M.~B., Chamberlin, P.~C., Eparvier, F.~G., {et~al.} 2017, Journal
  of Geophysical Research: Space Physics, 122, 2748,
  \dodoi{https://doi.org/10.1002/2016JA023512}

\bibitem[{{Uitenbroek}(2001)}]{Uitenbroek2001RH}
{Uitenbroek}, H. 2001, \apj, 557, 389, \dodoi{10.1086/321659}

\bibitem[{{Viereck} {et~al.}(2007){Viereck}, {Hanser}, {Wise}, {Guha}, {Jones},
  {McMullin}, {Plunket}, {Strickland}, \& {Evans}}]{Viereck2007EUVS}
{Viereck}, R., {Hanser}, F., {Wise}, J., {et~al.} 2007, in Society of
  Photo-Optical Instrumentation Engineers (SPIE) Conference Series, Vol. 6689,
  Solar Physics and Space Weather Instrumentation II, ed. S.~{Fineschi} \&
  R.~A. {Viereck}, 66890K, \dodoi{10.1117/12.734886}

\bibitem[{Watanabe(2014)}]{Watanabe2014SolarC}
Watanabe, T. 2014, in Space Telescopes and Instrumentation 2014: Optical,
  Infrared, and Millimeter Wave, ed. J.~M.~O. Jr., M.~Clampin, G.~G. Fazio, \&
  H.~A. MacEwen, Vol. 9143, International Society for Optics and Photonics
  (SPIE), 91431O, \dodoi{10.1117/12.2055366}

\bibitem[{{Wauters} {et~al.}(2022){Wauters}, {Dominique}, {Milligan},
  {Dammasch}, {Kretzschmar}, \& {Machol}}]{Wauters2022M67Lya}
{Wauters}, L., {Dominique}, M., {Milligan}, R., {et~al.} 2022, \solphys, 297,
  36, \dodoi{10.1007/s11207-022-01963-0}

\bibitem[{Woods {et~al.}(2006)Woods, Kopp, \&
  Chamberlin}]{Woods2006TSIVariation}
Woods, T.~N., Kopp, G., \& Chamberlin, P.~C. 2006, Journal of Geophysical
  Research: Space Physics, 111, \dodoi{https://doi.org/10.1029/2005JA011507}

\bibitem[{{Woods} {et~al.}(1993){Woods}, {Rottman}, \&
  {Ucker}}]{Rottman1993UARSPaper2}
{Woods}, T.~N., {Rottman}, G.~J., \& {Ucker}, G.~J. 1993, \jgr, 98, 10,679,
  \dodoi{10.1029/93JD00463}

\bibitem[{Woods {et~al.}(2004)Woods, Eparvier, Fontenla, Harder, Kopp,
  McClintock, Rottman, Smiley, \& Snow}]{Woods2004Lya20Percent}
Woods, T.~N., Eparvier, F.~G., Fontenla, J., {et~al.} 2004, Geophysical
  Research Letters, 31, \dodoi{https://doi.org/10.1029/2004GL019571}

\bibitem[{{Woods} {et~al.}(2012){Woods}, {Eparvier}, {Hock}, {Jones},
  {Woodraska}, {Judge}, {Didkovsky}, {Lean}, {Mariska}, {Warren}, {McMullin},
  {Chamberlin}, {Berthiaume}, {Bailey}, {Fuller-Rowell}, {Sojka}, {Tobiska}, \&
  {Viereck}}]{Woods2012EVE}
{Woods}, T.~N., {Eparvier}, F.~G., {Hock}, R., {et~al.} 2012, \solphys, 275,
  115, \dodoi{10.1007/s11207-009-9487-6}

\bibitem[{{Woods} {et~al.}(2017){Woods}, {Caspi}, {Chamberlin}, {Jones},
  {Kohnert}, {Mason}, {Moore}, {Palo}, {Rouleau}, {Solomon}, {Machol}, \&
  {Viereck}}]{Woods2017MinXSS}
{Woods}, T.~N., {Caspi}, A., {Chamberlin}, P.~C., {et~al.} 2017, \apj, 835,
  122, \dodoi{10.3847/1538-4357/835/2/122}

\bibitem[{{Wuelser} {et~al.}(2004){Wuelser}, {Lemen}, {Tarbell}, {Wolfson},
  {Cannon}, {Carpenter}, {Duncan}, {Gradwohl}, {Meyer}, {Moore}, {Navarro},
  {Pearson}, {Rossi}, {Springer}, {Howard}, {Moses}, {Newmark},
  {Delaboudiniere}, {Artzner}, {Auchere}, {Bougnet}, {Bouyries}, {Bridou},
  {Clotaire}, {Colas}, {Delmotte}, {Jerome}, {Lamare}, {Mercier}, {Mullot},
  {Ravet}, {Song}, {Bothmer}, \& {Deutsch}}]{Wuelser2004STEREO_A_EUVI}
{Wuelser}, J.-P., {Lemen}, J.~R., {Tarbell}, T.~D., {et~al.} 2004, in Society
  of Photo-Optical Instrumentation Engineers (SPIE) Conference Series, Vol.
  5171, Telescopes and Instrumentation for Solar Astrophysics, ed.
  S.~{Fineschi} \& M.~A. {Gummin}, 111--122, \dodoi{10.1117/12.506877}

\bibitem[{Yan {et~al.}(2022)Yan, Dang, Cao, Cui, Zhang, Liu, \&
  Lei}]{Yan2022EarthVenusMarsSolarFlares}
Yan, M., Dang, T., Cao, Y.-T., {et~al.} 2022, The Astrophysical Journal, 939,
  23, \dodoi{10.3847/1538-4357/ac92ff}

\bibitem[{{Zarro} \& {Lemen}(1988)}]{Zarro1988CondEvap}
{Zarro}, D.~M., \& {Lemen}, J.~R. 1988, \apj, 329, 456, \dodoi{10.1086/166391}

\bibitem[{{Zhang} {et~al.}(2019){Zhang}, {Chen}, {Wu}, {Chang}, {Hu}, {Su},
  {Zhang}, {Wang}, {Liang}, {Ma}, {Guo}, {Cai}, {Zhang}, {Huang}, {Peng},
  {Tang}, {Zhao}, {Zhou}, {Wang}, {Song}, {Ma}, {Xu}, {Yang}, {Lu}, {He},
  {Tao}, {Ma}, {Lv}, {Bai}, {Cao}, {Huang}, \& {Gan}}]{Zhang2019HXI}
{Zhang}, Z., {Chen}, D.-Y., {Wu}, J., {et~al.} 2019, Research in Astronomy and
  Astrophysics, 19, 160, \dodoi{10.1088/1674-4527/19/11/160}

\end{thebibliography}
\bibliographystyle{aasjournal}

\end{document}